\documentclass[prl,twocolumn,aps,superscriptaddress,longbibliography]{revtex4-1}

\usepackage{bm}%
\usepackage{float}
\expandafter\ifx\csname package@font\endcsname\relax\else
 \expandafter\expandafter
 \expandafter\usepackage
 \expandafter\expandafter
 \expandafter{\csname package@font\endcsname}%
\fi

\usepackage{array}
\usepackage{amsmath,amssymb}
\usepackage{MnSymbol}
\usepackage{tikz-feynman,contour} 
\usetikzlibrary{shapes,arrows,positioning,automata,backgrounds,calc,er,patterns}
\usepackage{feynmf}
\tikzfeynmanset{compat=1.0.0} 

\usepackage{graphicx}
\usepackage{mathrsfs}
\usepackage{bm}
\usepackage{mathtools}
\usepackage{epstopdf}
\usepackage{hyperref}

\hypersetup{%
   pdfpagemode=None, 
   pdfstartpage=1,
   pdfmenubar=true,
   pdftoolbar=true,
   colorlinks = true,
   linkcolor=blue,
   citecolor=blue,
   urlcolor=blue,
   bookmarksopen=false
 }
 \usepackage{amsmath}
\usepackage{amsfonts}
\usepackage{mathtools}
\usepackage{graphicx}
\usepackage{fixme}
\usepackage{color}

\begin{document}

\title{Charged polarons and molecules in a Bose-Einstein Condensate}




\author{Esben Rohan Christensen}
\affiliation{Center for Complex Quantum Systems, Department of Physics and Astronomy, Aarhus University, Ny Munkegade 120, DK-8000 Aarhus C, Denmark.}
\author{Arturo Camacho-Guardian}
\affiliation{T.C.M. Group, Cavendish Laboratory, University of Cambridge, JJ Thomson Avenue, Cambridge, CB3 0HE, U.K.}
\affiliation{Center for Complex Quantum Systems, Department of Physics and Astronomy, Aarhus University, Ny Munkegade 120, DK-8000 Aarhus C, Denmark.}
\author{Georg M. Bruun} 
\affiliation{Center for Complex Quantum Systems, Department of Physics and Astronomy, Aarhus University, Ny Munkegade 120, DK-8000 Aarhus C, Denmark.}
\affiliation{Shenzhen Institute for Quantum Science and Engineering and Department of Physics, Southern University of Science and Technology, Shenzhen 518055, China. }

\begin{abstract}
We investigate the properties a mobile ion immersed in a Bose-Einstein condensate (BEC) using different theoretical approaches. A coherent state variational ansatz predicts  that 
the ion spectral function exhibits several branches in addition to polaronic quasiparticle states, and we employ a diagrammatic analysis of the ion-atom scattering in the BEC to identify them as 
arising from the binding of an increasing number of bosons to the ion. We develop a simplified model describing the formation of 
these molecular ions showing that their spectral weight 
scales with the number of bound atoms.  The number of atoms in the dressing cloud around the ion are calculated from thermodynamic arguments, and we 
 finally show that the  dynamics ensuing the injection of an ion into the BEC  exhibits various regimes governed by coherent quasiparticle 
 propagation and decay. 
 \end{abstract}

\date{\today}
\maketitle
The versatility and control of  atomic gases  make them powerful platforms for quantum simulation of  many-body
 systems~\cite{Bloch2008,Bloch2012}. Ions immersed in atomic gases represent an exciting new research  direction due to their 
 hybrid nature, which  enables new functionalities and  broader simulation capabilities. In particular, the excellent control  of the motional and 
 internal degrees of individual ions opens up new opportunities to explore the interaction between a small quantum system and its environment, and to  address fundamental 
 questions regarding  cooling, decoherence, and entanglement. The ion can also act as a local probe, which  has indeed already been exploited in classic  
 experiments investigating vortices~\cite{Yarmchuk1979} and the properties of superfluid liquid $^4$He~\cite{Meyer1958,Atkins1959,Gross1962} and
   $^3$He~\cite{Ahonen1976,Roach1977,Ahonen1978,Salomaa1980,Baym1979}. 

 Experiments on ions in atomic gases have explored atom-ion collisions,  sympathetic cooling, controlled 
 chemistry~\cite{Grier2009,Zipkes2010,Harter2012,Ratschbacher2012,Kleinbach2018,Sikorsky2018,Feldker2020,Schmidt2020},  transport~\cite{dieterle2020}, and molecular 
 formation~\cite{Dieterle2020b}. Theoretically, 
 the Fr\"ohlich model, valid for weak ion-atom interaction, was used to explore an
 ion in an atomic Bose-Einstein condensate (BEC)~\cite{Casteels2011} and three-body recombination  dynamics was studied in Refs.~\cite{Gao2010,Krukow2016}. 
 Several papers have predicted the formation of  molecular ions
based on kinetic and mean-field approaches~\cite{Lukin2002,Massignan2005}, quantum defect theory~\cite{Gao2010}, 
 and time-dependent Hartree and Monte-Carlo  calculations~\cite{Schurer2017,astrakharchik2020}. 
 
 Inspired by this exciting development, we investigate here a mobile ion immersed in a BEC. Using a variational ansatz allowing for the dressing of an infinite number of Bogoliubov modes, 
 we show that the ion spectral functions has several branches. A diagrammatic analysis of the ion-atom scattering shows that they arise from the binding of an increasing number of atoms to the ion.
Inspired by this, we develop a simple model for the formation of these molecular ions and show that their spectral weight is proportional to the number of atoms bound to the ion. We use thermodynamic
arguments to calculate the number of atoms in the dressing cloud around the ion, and  finally demonstrate how the quantum dynamics  after an ion is injected into the BEC is characterised by coherent evolution  and  decay into the molecules.

\paragraph{Model.-}
 Consider an ion of mass $m$ immersed in a  BEC of atoms of mass $m_B$. The Hamiltonian is
  \begin{align}
\hat{H} =& \sum_{\mathbf{k}} \left(\frac{\mathbf{k}^2}{2m} \hat{a}^\dagger_\mathbf{k} \hat{a}_\mathbf{k} +  \frac{\mathbf{k}^2}{2m_B}\hat{b}^\dagger_\mathbf{k} \hat{b}_\mathbf{k}\right) + \frac{g_{\text{B}}}{2} \sum_{\mathbf{k},\mathbf{k'},\mathbf{q}} \hat{b}^\dagger_\mathbf{k+q} \hat{b}^\dagger_\mathbf{k'-q}\hat{b}_\mathbf{k'} \hat{b}_\mathbf{k} \nonumber\\
&+ \sum_{\mathbf{k}, \mathbf{k'},\mathbf{q}} V(\mathbf{q}) \hat{a}^\dagger_{\mathbf{k'-q}} \hat{a}_{\mathbf{k'}} \hat{b}^\dagger_{\mathbf{k}+\mathbf{q}}
\hat{b}_\mathbf{k},
\label{H0}
\end{align}
where $\hat{a}^\dagger_\mathbf{k}$ and $\hat{b}^\dagger_\mathbf{k}$ creates an ion and a boson respectively with momentum ${\mathbf k}$. 
We describe the BEC of density $n_0$ using Bogoliubov theory giving the dispersion 
$E_\mathbf{k}=\sqrt{\epsilon_\mathbf{k}^2+2n_0 g_{B} \epsilon_\mathbf{k}}$ with $\epsilon_\mathbf{k}=\mathbf k^2/2m_B$ and  $g_{B}=4\pi a_{B}/m_B$
with   $a_B$ the atom-atom  scattering length. The atom-ion interaction is $V({\mathbf k})$,  and we use units where 
the system volume and $\hbar$ are unity.

\begin{figure}[htb]
    \includegraphics[width=\columnwidth,]{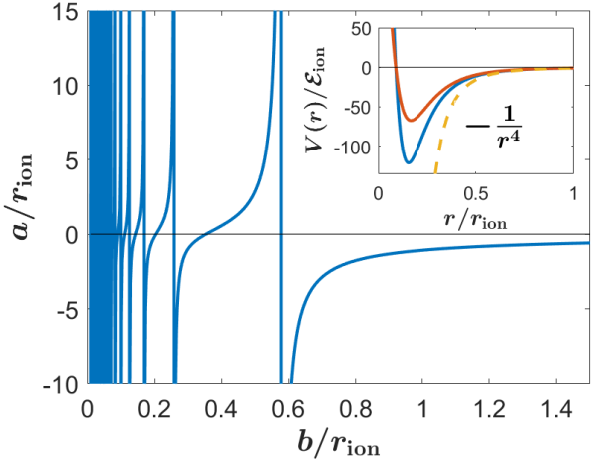}
    \caption{The atom-ion $s$-wave scattering length  $a$ as a function of $b$. The inset shows the 
    atom-ion potential for $b/r_\text{ion}=0.3$ and 0.35.}
    \label{scatlen}
\end{figure}

In real space, the atom-ion interaction has the long-range asymptotic form $V({\mathbf r})\sim -\alpha/r^4$, where $\alpha$ is proportional to the polarisability of the 
atoms~\cite{Tomza2019}. A characteristic length scale of the interaction is therefore $r_\text{ion}=\sqrt{2m_r\alpha}$ with $m_r^{-1}=m^{-1}+m_B^{-1}$, 
and using the polarisability of atoms like $^{87}$Rb and 
$^{23}$Na this gives $r_\text{ion}\sim{\mathcal O}(10^2)$nm~\cite{Massignan2005}. This  is of the same order as the average interparticle distance for a typical
 BEC with density $n_0\sim10^{14}$cm$^{-3}$, and it is therefore crucial to include the asymptotic  form of $V({\mathbf r})$ in our analysis. To do this,  
we use the effective interaction~\cite{Krych2015}
\begin{equation}
    V(r) = -\frac{\alpha}{(r^2+b^2)^2}\frac{r^2-c^2}{r^2+c^2},
    \label{potential}
\end{equation}
where the parameter $c$ establishes a repulsive barrier such that the potential is repulsive (attractive) for $c<r (c>r)$, while $b$ is related to the depth of the potential. 
We have $V(0)=\alpha/b^4$, which is large compared to any other relevant energy in 
order to mimic the strong repulsion when the electron clouds of the atom and the ion  overlap. In the inset of Fig.~\ref{scatlen}, we plot 
$V(r)$ in units of $\mathcal{E}_\text{ion}=1/2m_rr_\text{ion}^2$ for two different values of $b$. Here and in the rest of the Letter, we take  $c=0.0023r_\text{ion}$, which is experimentally relevant for $^{87}$Rb where the potential is repulsive at $r\lesssim 10a_0$. Also, we use $m=m_B$.

\paragraph{Two-body physics.-} 
In Fig.~\ref{scatlen}, the  atom-ion scattering length $a$, obtained by solving the zero energy $s$-wave Schr\"odinger equation with the potential  $V(r)$, is 
plotted as a function of $b$. It exhibits several divergencies, which correspond to the 
 emergence of two-body bound states. The first bound state appears for $b/r_\text{ion}\simeq 0.58,$ and more bound states appear as the 
 atom-ion potential becomes deeper with decreasing $b$. 
 As we shall now see, the long-range nature of the atom-ion interaction and the presence of several two-body bound states give rise to the presence of several quasiparticle and mesoscopic molecular states in the corresponding many-body problem.

 \paragraph{Variational ansatz.-}
 To analyse this challenging interplay between few- and many-body physics, we employ set of a different theoretical techniques. Anticipating 
the formation of molecular ions involving many bosons bound to ion, we first employ a time-dependent coherent state variational ansatz. 
  In the frame co-moving with the ion, obtained by  the unitary transformation $\hat{S} = \exp(-i \hat{\mathbf{r}}\cdot \hat{\mathbf{P}}_B)$ where 
  $\hat{\mathbf r}$ is the position of the ion and $\hat{\mathbf P}_B=\sum_{\mathbf{k}}\mathbf{k}
 \hat \beta^\dagger_{\mathbf{k}} \hat \beta_{\mathbf{k}}$ is the total momentum of the atoms~\cite{Lee1953}, 
   it reads~\cite{Shashi2014,Shchadilova2016}
\begin{gather}
    |\Psi (t) \rangle = e^{-i \phi (t) } e^{\sum_\mathbf{k} [\gamma_\mathbf{k}(t) \hat\beta^\dagger_\mathbf{k} - \gamma^\ast_\mathbf{k}(t) 
    \hat\beta_\mathbf{k}]}  |\Psi (0)\rangle.
    \label{cansatz}  
\end{gather}
Here, the operator $\hat\beta^\dagger_\mathbf{k}=u_{\mathbf k}\hat b^\dagger_{\mathbf k}+v_{\mathbf k}\hat b_{-\mathbf k}$ creates a 
Bogoliubov mode with momentum ${\mathbf k}$ and energy $E_{\mathbf k}$. 
The initial state $ |\Psi (0)\rangle=\hat a_{\mathbf{k}=0}^\dagger|\text{BEC}\rangle$ corresponds to the injection of a zero momentum ion in the BEC.

The Euler-Lagrange equations give  the following equations of motion for the parameters $\phi$ and
$\gamma_\mathbf{k}$~\cite{SM}
\begin{gather}
  \dot{\phi}=-\frac{P_B^2}{2m} + n_0 V(\mathbf{0}) + \frac{\sqrt{n_0}}{2}\sum_\mathbf{k} V(\mathbf{k})W_\mathbf{k}(\gamma^\ast_\mathbf{k}+\gamma_\mathbf{k}),
  \label{phieqn}\\ \nonumber
i\dot{\gamma}_\mathbf{k} = \sqrt{n_0}V(\mathbf{k}) W_\mathbf{k} + \left( E_\mathbf{k}+\frac{\mathbf{k}^2}{2m}+\mathbf{k}\cdot\frac{\mathbf{P}_B}{m} \right) \gamma_\mathbf{k} \\ 
+\sum_{\mathbf{k'}}  V(\mathbf{k}'-\mathbf{k}) C_{\mathbf{k}\mathbf{k'}} \gamma_\mathbf{k'}
 - \sum_{\mathbf{k'}}  V(\mathbf{k}'+\mathbf{k})\tilde C_{\mathbf{k}\mathbf{k'}} \gamma_{\mathbf{k'}}^\ast\label{betapoteq1},
\end{gather} 
where ${\mathbf P}_B=\sum_{\mathbf k}|\gamma_{\mathbf k}|^2{\mathbf k}$, $W_\mathbf{k}=\sqrt{\epsilon_\mathbf{k}/E_\mathbf{k}}$,
$C_{\mathbf{k}\mathbf{k'}}=u_{\mathbf{k}} u_{\mathbf{k}'} + v_{\mathbf{k}} v_{\mathbf{k}}'$,  
$\tilde C_{\mathbf{k}\mathbf{k}'}= u_\mathbf{k}v_{\mathbf{k'}}+v_\mathbf{k}u_\mathbf{k'}$, and 
 $u_\mathbf{k}^2/v_\mathbf{k}^2 = [ (\epsilon_\mathbf{k}+g_{B}n_0)/E_\mathbf{k}\pm1 ]/2$  
the usual coherence factors.

We solve Eqs.~\eqref{phieqn}-\eqref{betapoteq1} numerically, from which  the dynamical overlap
  $S(t)=\langle \Psi(0) | \Psi(t) \rangle = e^{-i \phi(t)} e^{-\frac{1}{2} \sum_\mathbf{k}{|\gamma_{\mathbf{k}}|}^2}$ with the initial state can be calculated. The  
  impurity spectral function  can then  be obtained by a Fourier transform
   $ A(\omega)=\text{Re} \int_0^\infty S(t) e^{i \omega t} dt/\pi$~\cite{Knap2012,SM}.

\paragraph{Dressing cloud.-}
To further analyse the nature of the many-body states, we use a thermodynamic argument to calculate the 
  number of atoms $\Delta N$ in the dressing cloud around the ion. This  gives~\cite{Massignan2005,Massignan2011}
  \begin{align}
    \Delta N =-{\left( \frac{\partial \mu_I}{\partial n_0} \right)}{\left( \frac{\partial n_0}{\partial \mu_B} \right)}_{n_I=0}= - {\left( \frac{\partial \mu_I}{\partial \mu_B} \right)}_{n_I=0},
    \label{DeltaN1}
\end{align}
where $\mu_I$ is the energy change when the ion is added to the BEC, $\mu_B=g_Bn_0$ is the chemical potential of the atoms, and $n_I$ is the ion 
density, which is zero for  a single ion. For a given many-body state 
with energy $\mathcal{E}_j$, we set $\mu_I=\mathcal{E}_j$ in Eq.~\eqref{DeltaN1} to calculate $\Delta N$.

\begin{figure}[htb]
 \includegraphics[width=\columnwidth]{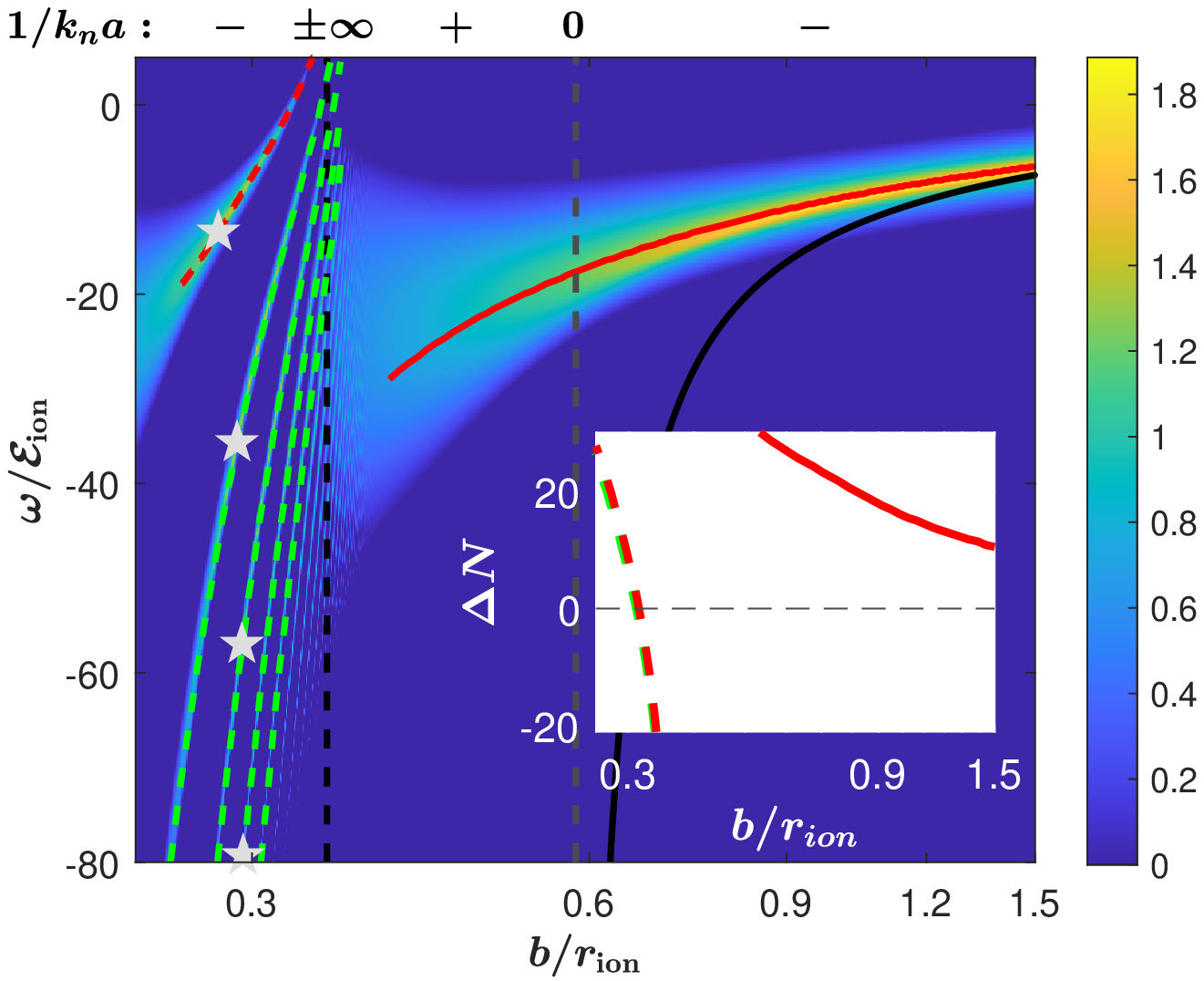}   
 \includegraphics[width=\columnwidth]{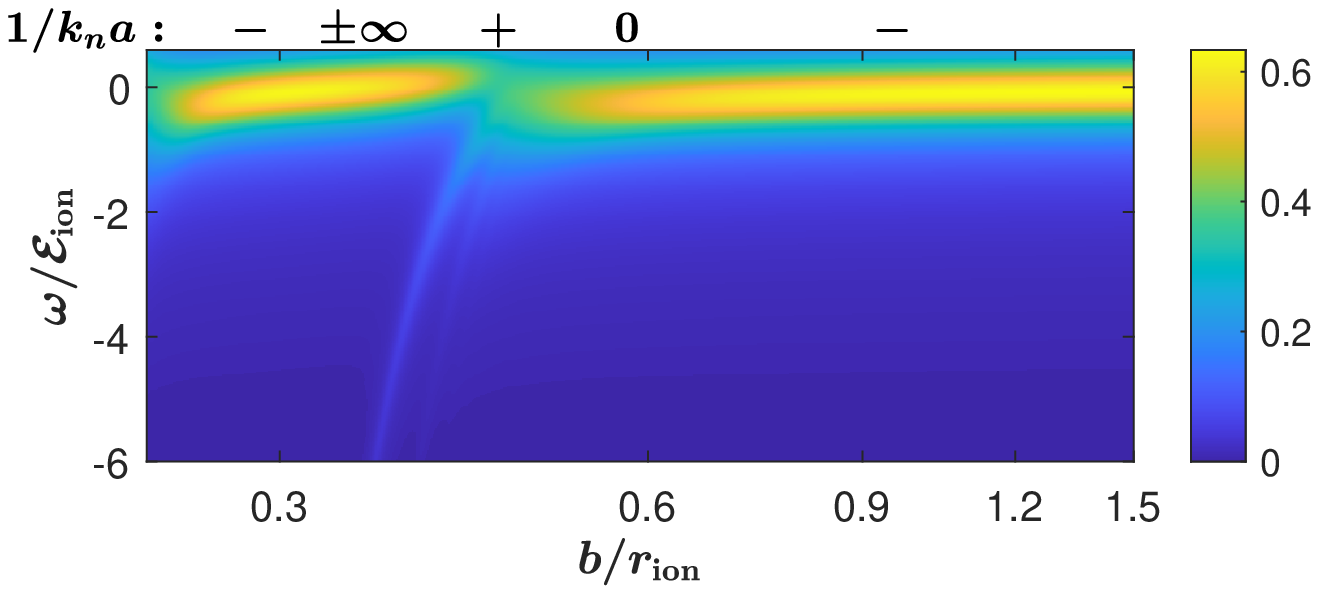}
\caption{The zero momentum ion spectral function $A(\omega)$ as function of the potential parameter $b$ and 
the corresponding scattering length $a$ for  $n_0r_\text{ion}^3=1$(top) and  $n_0r_\text{ion}^3=0.01$ (bottom). The black dashed line is the mean-field energy,
 the  red line  is the ladder approximation for the  attractive polaron present for $b/r_\text{ion}\gtrsim0.5$, the red dashed line the 
  attractive polaron present for $b/r_\text{ion}\lesssim0.34$, and the  green lines are the molecular states obtained from the Bethe-Salpeter equation.  The stars $\star$ 
  signify the breakdown of the Bogoliubov approximation. The number of 
  atoms in the dressing cloud is shown in the inset with the same color coding as the main figure.
 }
    \label{comparenobnd1}
\end{figure}

\paragraph{Polarons.-}
The ion spectral function 
obtained from the variational ansatz is shown in Fig.~\ref{comparenobnd1}(top) for a density $n_0r_\text{ion}^3=1$ and zero temperature as a function 
of $b$ and  the corresponding scattering length $a$. 
 For large $b$ meaning weak coupling  $1/k_na\ll-1$ with $k_n^3/6\pi^2=n_0$,   there is a well-defined 
quasiparticle  with mean-field energy $E=2\pi a n_0/m_r$. Its energy decreases with decreasing $b$
 (increasing $1/k_na$) corresponding to an increasing depth of the potential, and the mean-field expression eventually 
breaks down. This quasiparticle  is the attractive Bose polaron for the ion in direct analogy with what is observed for neutral 
impurities~\cite{Jorgensen2016,Hu2016,Ardila2019,Yan2020}. In the inset of Fig.~\ref{comparenobnd1}, we see that number of bosons 
in the dressing cloud around the ion can be quite large reflecting the strength and range of the atom-ion interaction.
In the weak coupling limit $b/r_{ion}\gg 1$, we recover the mean-field result $\Delta N=-a/a_B$~\cite{Massignan2005}. The attractive polaron remains a stable ground state with 
decreasing $b$ but with a very small residue.
Since we have added a small imaginary part to the frequency for numerical reasons, its quasiparticle peak becomes indistinguishable from the many-body continuum starting 
at energies just above~\cite{footnote}. 

Figure \ref{comparenobnd1} shows a number of  new states that emerge in the regime  $b/r_{ion} <0.58$ where the atom-ion interaction supports a two-body bound state. 
We have  $a>0$ for $0.58>b/r_{ion} >0.35$ and $a\le 0$ for  $0.35 \ge b/r_{ion} > 0.26$
where another bound state emerges, see Fig.~\ref{scatlen}. Consider first the branch with the highest energy  emerging for $b/r_{ion} \simeq0.34\Rightarrow 1/k_n a\simeq -1.45$, highlighted by a red dashed line.  
Its energy $\varepsilon_{P}$ is larger than zero for $b/r_{ion} \gtrsim 0.32$ where the number of particles $\Delta N$ in its dressing cloud is negative as shown in the inset.
From this we conclude that it is a repulsive polaron quasiparticle. Its energy becomes negative for  $b/r_{ion} \lesssim 0.32$ where $\Delta N>0$ showing that it smoothly 
evolves into an attractive polaron with increasing depth of the ion-atom interaction potential. Note that there is no analogue of such an attractive polaron
 when there is a bound state  for a short-range interaction. 
 
\paragraph{Molecular ions.-}
  Furthermore, Fig.~\ref{comparenobnd1}  shows several new low energy states emerging together with the repulsive polaron at $b/r_{ion} \lesssim 0.34$. As we will now 
  demonstrate, they arise from the binding of  $1,2,\ldots$ bosons to the ion. Consider the  
  scattering matrix between an ion with momentum/energy $k_1=(\mathbf{k}_1,i\omega_1)$ and an atom with momentum/energy $k_1=(\mathbf{k}_1,i\omega_1)$.
  In the ladder approximation, it obeys the  Bethe-Salpeter equation~\cite{SM} 
\begin{gather}
\Gamma({\mathbf k}_1,{\mathbf  k}_2,{\mathbf q};i\omega_1+i\omega_2)=V({\mathbf q})-\sum_{q'}V({\mathbf q}')G_{11}({ k}_2-q')\nonumber \\
\times G({ k}_1+q')\Gamma(\mathbf{ k}_1+\mathbf{q}',{\mathbf  k}_2-\mathbf{q}',\mathbf{q-q}';i\omega_1+i\omega_2),
\label{BetheS1}
\end{gather}
where ${\mathbf q}$ is the momentum transfer, $G({k})=1/(i\omega-{\mathbf k}^2/2m)$ is the ion Green's function, 
  and  $G_{11}(k)=u^2_{\mathbf k}/(i\omega-E_{\mathbf k})-v^2_{\mathbf k}/(i\omega+E_{\mathbf k})$ is the normal BEC Green's function for the atoms. 
  The sum $\sum_{q'}\equiv T\sum_{i\omega}\int d^3q/(2\pi)^3$  is both over momenta $\mathbf q$ and Matsubara 
frequencies $i\omega$, and we analytically continue $i\omega\rightarrow \omega+i0_+$  as usual. Due to the long range of the atom-ion potential, it is essential to retain
its full momentum dependence  in Eq.~\eqref{BetheS1},  in contrast to the usual
 case of a short-range interaction between neutral atoms. 

The ion self-energy $\Sigma({\mathbf k},\omega)=n_0\Gamma({\mathbf k},0,0;\omega)$ describes the scattering of a single atom out of the BEC, 
and the quasiparticle energy is obtained by solving  $\varepsilon_{P,\mathbf k}=k^2/2m+\Sigma({\mathbf k},\varepsilon_{P,\mathbf k})$. The resulting ladder approximation 
has successfully  been applied to explain experimental results for neutral impurities in a BEC forming Bose polarons~\cite{Rath2013,Jorgensen2016,Hu2016,Ardila2019,Yan2020}. In   
the present case it yields the red line in Fig.~\ref{comparenobnd1}, which agrees very well with the variational result for the attractive polaron stable, whereas it fails to 
capture the lower lying states.

This can however be addressed by noting that a possible pole of the zero momentum scattering matrix gives the energy of a bound  state. Thus, by replacing  in Eq.~\eqref{BetheS1} 
the bare ion Green's function  with the polaron Green's function $G_j({k})=1/(i\omega-\varepsilon_P-{\mathbf k}^2/2m)$ 
will give the energy of a possible  dimer consisting of an atom bound to the polaron. This yields the top green line  in Fig.~\ref{comparenobnd1}. The excellent 
agreement with the variational ansatz show that this state indeed arises from the binding of an atom to the ion. 
We perform this procedure recursively by calculating the scattering matrix between this new molecular state 
and an atom, which then yields the second green line below the attractive polaron in  Fig.~\ref{comparenobnd1} and so on. 
Since energies obtained from this procedure agree very well with those  from the variational ansatz, we conclude that  
 these branches involve  the binding of one, two, $\ldots$ atoms to the ion. In the following, we refer for brevity to these states as molecular ions although they do have a non-zero 
 quasiparticle residue as is evident from Fig.~\ref{comparenobnd1}. We note that dimer states consisting of one atom bound to the ion have recently been observed~\cite{Dieterle2020b}, and 
 our prediction of molecular states involving more atoms is consistent with earlier results based on  different methods~\cite{Lukin2002,Massignan2005,Schurer2017,astrakharchik2020}.

 Note that these molecules are stable only for    
 $b$ significantly smaller than $b/r_{ion} =0.58$ where the two-body atom-ion  state emerges. 
 Hence,  many-body effects destabilize the binding of atoms to the ion as compared to the vacuum case. The molecules are 
 stable for $a>0$ and $a<0$ as opposed to the case of a short-range interaction, where similar states are predicted to exist only for $a<0$~\cite{Shchadilova2016}.

 The binding of additional atoms to the ion will eventually be halted by the repulsion between them giving a positive energy  $\sim a_B\Delta N^2$. 
 While this effect is not included in our theory, we can estimate when it becomes important by calculating the gas factor of the dressing cloud $\sqrt{n_\text{cl}a_B^3}$. 
 Here, $n_\text{cl}=\Delta N/\bar r^3$ is the average density of atoms in the dressing cloud with  $\bar r=[\int\!d^3rr^2|\phi({\mathbf r})|^2]^{1/2}$  the spatial size of the
  molecule with wave function $\phi({\mathbf r})$. As explicitly shown in the Supp.\ Mat.~\cite{SM}, the size of the  molecular states is $\sim r_\text{ion}$ and decreases as 
they  becomes increasingly bound.  The  $\star$'s in Fig.~\ref{comparenobnd1} indicate when the gas factor of a given molecular state becomes larger than $0.1$.
 A reliable description of this region requires one to go beyond Bogoliubov theory.



\paragraph{Simplified model.-}    
The basic physics of the  binding of bosons  to the ion can be described using the Hamiltonian 
\begin{align}
\hat{H}_s=\sum_{l=0}^\infty\left\{[\varepsilon_P+\varepsilon_B(l-1)]\hat c_l^\dagger\hat c_l+g\sqrt{n_0}\hat c_{l+1}^\dagger\hat c_{l}+\text{h.c.}\right\}.
\end{align}
Here, $\hat c_l$ creates a state with $l$  bosons bound the polaron, $\varepsilon_B<0$ is the energy released by the binding of a boson, and $g\sqrt{n_0}$ is the matrix element
 for this process. Note that this is proportional to $\sqrt{n_0}$ since the boson is taken from the BEC with density $n_0$. 
 This also means that we can suppress the momentum 
 since this is zero for all states. The model is easily solved giving a continued fraction form of the zero momentum ion Green's function 
\begin{align}
G(\omega)^{-1}=\omega-\varepsilon_P-\cfrac{g^2n_0}{\omega-\varepsilon_B-\cfrac{g^2n_0}{\omega-2\varepsilon_B-\cfrac{g^2n_0}{\omega-3\varepsilon_B-\ldots}}}.
\end{align}
For $g^2n_0/\varepsilon_B^2\ll 1$, the highest energy pole is  
$\simeq \varepsilon_P$ corresponding to the repulsive polaron and there is  an infinite ladder of poles with energies $\simeq \varepsilon_P-l\varepsilon_B$ 
 corresponding to states with $l=1,2,\ldots $ bosons bound to the ion. The residue of these states is  $(g^2n_0/\varepsilon_B^2)^l\propto n_0^l$
 reflecting that they involve $l$ bosons taken from the BEC. This scaling explains the decreasing spectral weight of the deeper molecular
  lines seen  Fig.~\ref{comparenobnd1}. 
  
  It also means that the relative spectral weight of the different lines  depends on the BEC density. This is illustrated in 
  Fig.~\ref{comparenobnd1}(bottom), which shows the ion spectral function for $n_0r_\text{ion}^3=0.01$. We see that only two states with significant spectral 
  weight emerge for  $b/r_{ion} <0.58$  when the atom-ion potential supports a bound state:  The new polaron and the  highest molecular state with one boson bound to the ion. 
Since the ground state remains the attractive polaron, this is consistent with the finding that for a static ion in the dilute limit, there are 
  $2\nu_s+1$ solutions to the Gross-Pitaevskii equation where $\nu_s$ is the number of two-body bound states of the atom-ion interaction 
  potential~\cite{Massignan2005,PietroPrivate}. 
The small spectral weight of the bound states involving more than one boson also means that they are quite sensitive to additional damping.

\begin{figure}
\includegraphics[width=0.49\columnwidth]{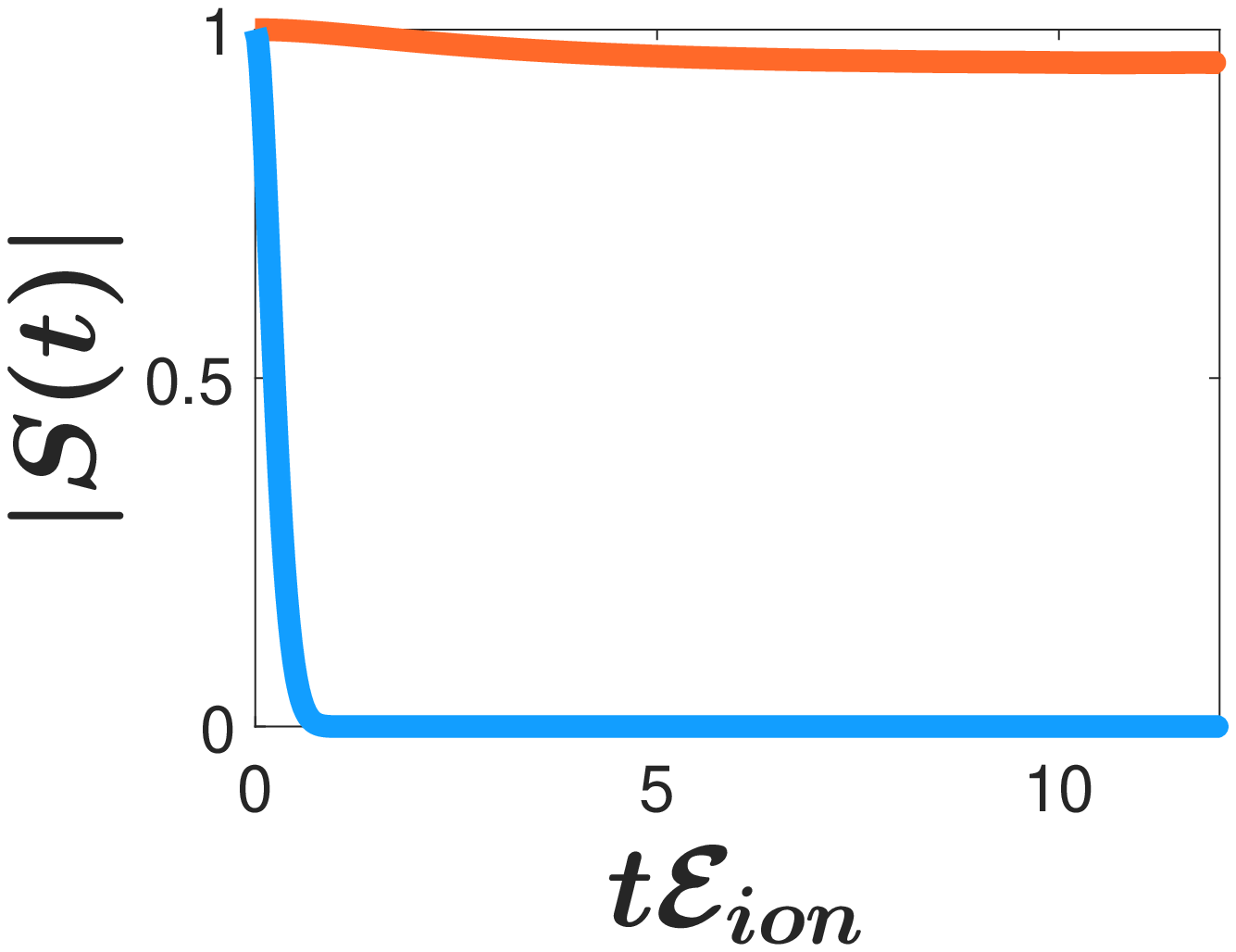}
\includegraphics[width=0.49\columnwidth]{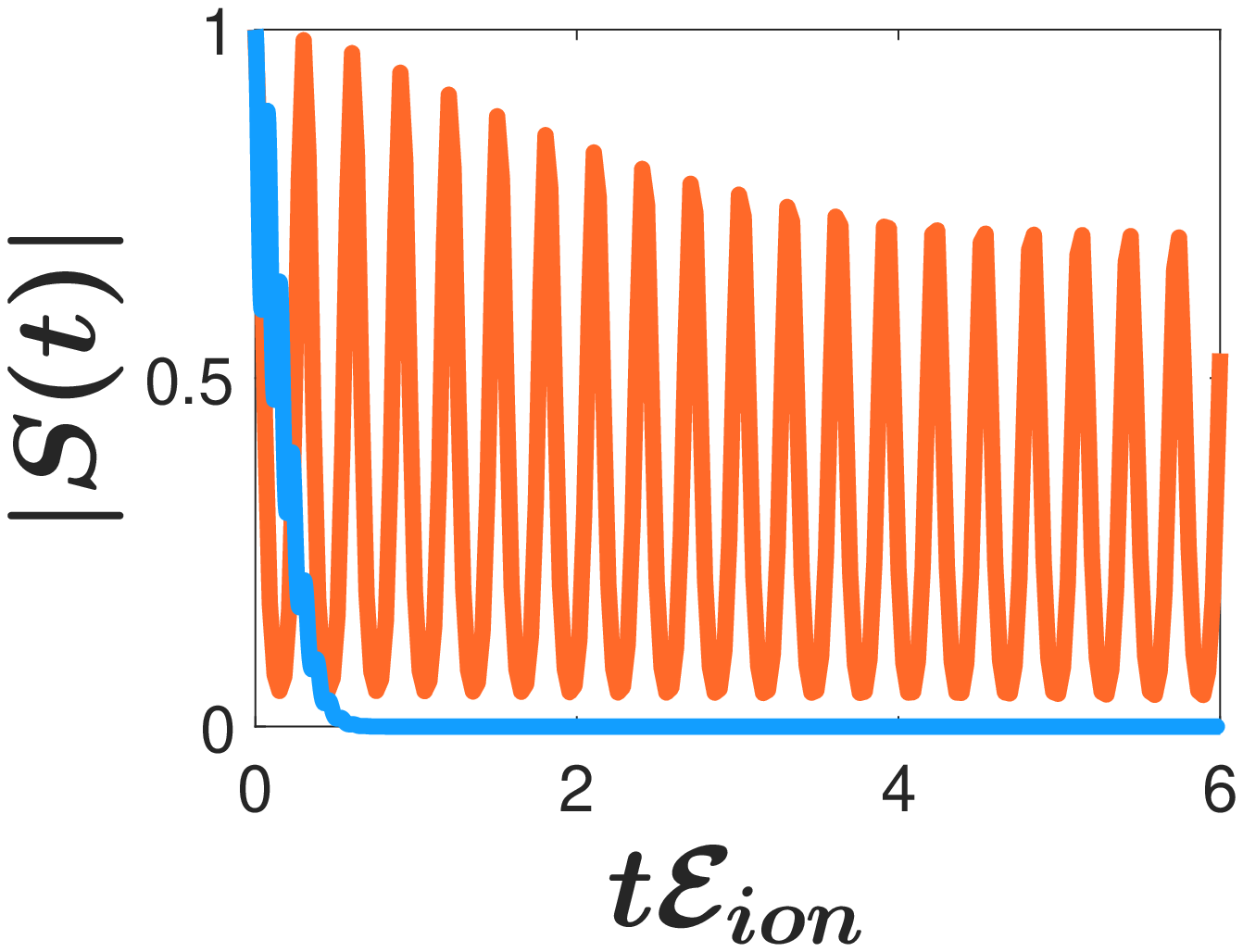} 
\caption{In the left panel $|S(t)|$ is shown for $b/r_{ion}=2$ (orange) and $b/r_{ion}=0.5$ (blue). The right panel shows $|S(t)|$ for $b/r_{ion}=0.3$ (orange) and $b/r_{ion}=0.25$ (blue). }
\label{spectral1}
\end{figure}

\paragraph{Dynamics.-} We finally investigate the quantum dynamics after a zero momentum ion is injected in the  BEC. The 
dynamical overlap $S(t)=\langle \Psi(0) | \Psi(t) \rangle $ 
 is plotted in Fig.~\ref{spectral1}.  For $b/r_{ion}=2$, we have $|S(t)|\rightarrow Z$ for $t\rightarrow \infty$ where $Z$ 
is the quasiparticle residue of the attractive polaron~\cite{Shchadilova2016,Nielsen2019}. For $b/r_{ion}=0.5$ on the other hand, 
 $S(t)$  decreases monotonically to zero since there is no well-defined quasiparticle, see Fig.~\ref{comparenobnd1}. 
 In the right panel of Fig.~\ref{spectral1}, we plot $S(t)$ when the molecular states are present. For $b/r_{ion}=0.3$ (orange), 
  $S(t)$ oscillates with an almost constant amplitude after an initial decay. These oscillations arise from a coherent population of the molecular states and the  
  polaron, see  Fig.~\ref{comparenobnd1}.
  For $b/r_{ion}=0.25$ (blue) on the other hand, the  polaron is strongly 
damped giving rise to decoherence and $S(t)$ therefore decays monotonically to zero, see Fig. \ref{comparenobnd1}.  
    
Figure   \ref{spectral1} shows that the many-body time-scale   is $\tau_{\text{ion}}\approx1/\mathcal E_{\text{ion}}= 13.5\mu\text{s}$ 
for $r_{\text{ion}}=100\text{nm}$. This should be compared to the  three-body recombination time $\tau_\text{3B}=1/K_3n_0^2$. 
Taking  $K_3\approx 3.3-6\times 10^{25}\text{cm}^{6}/\text{s}$~\cite{Harter2021,dieterle2020} and 
    $n_0=10^{14}\text{cm}^{-3}$ for a typical BEC yields  $\tau_\text{3B}\approx 160- 300\mu\text{s}$, showing that the many-body phenomena described here should be observable 
    before three-body decay sets in.

  

{\it Conclusions and outlook.-} 
Using several theoretical methods, we  studied the static and dynamical properties of a mobile  ion in a BEC. 
 The long-range nature of the atom-ion interaction was shown to result in a rich spectrum with 
several quasiparticle and  molecular states. We demonstrated that the quantum dynamics
 after a quench  where the ion is injected into the BEC is characterised by coherent oscillations between the different states as well as decay.  
 Our work demonstrates  the diverse  and exciting physics that can be realised in   ion-atom systems and
 motivates future investigations into these hybrid systems.  In particular,  dimer states consisting of one atom bound to the ion have recently been observed, and it would be very interesting to 
 extend this experimental search to the predicted deeper lying larger molecular ions preferably using a high density BEC~\cite{Dieterle2020b}. Also, radio-frequency and Ramsey spectroscopy have been used to measure the spectral function and 
 the  dynamics for neutral  impurities in a BEC~\cite{Jorgensen2016,Hu2016,Ardila2019,Yan2020,Skou2020}, and analogous probes for charged 
 impurities should be highly useful.

{\it Acknowledgments.-} We acknowledge financial support from the Villum Foundation, the Independent Research Fund Denmark-Natural Sciences via Grant No. DFF -8021-00233B, and US Army CCDC Atlantic Basic and Applied
Research via grant W911NF-19-1-0403. We thank P.\ Massignan for very useful comments, and 
T. Pohl and K. M\o lmer for  discussions.

\bibliography{Ion_Polaron}

\begin{thebibliography}{48}%
\makeatletter
\providecommand \@ifxundefined [1]{%
 \@ifx{#1\undefined}
}%
\providecommand \@ifnum [1]{%
 \ifnum #1\expandafter \@firstoftwo
 \else \expandafter \@secondoftwo
 \fi
}%
\providecommand \@ifx [1]{%
 \ifx #1\expandafter \@firstoftwo
 \else \expandafter \@secondoftwo
 \fi
}%
\providecommand \natexlab [1]{#1}%
\providecommand \enquote  [1]{``#1''}%
\providecommand \bibnamefont  [1]{#1}%
\providecommand \bibfnamefont [1]{#1}%
\providecommand \citenamefont [1]{#1}%
\providecommand \href@noop [0]{\@secondoftwo}%
\providecommand \href [0]{\begingroup \@sanitize@url \@href}%
\providecommand \@href[1]{\@@startlink{#1}\@@href}%
\providecommand \@@href[1]{\endgroup#1\@@endlink}%
\providecommand \@sanitize@url [0]{\catcode `\\12\catcode `\$12\catcode
  `\&12\catcode `\#12\catcode `\^12\catcode `\_12\catcode `\%12\relax}%
\providecommand \@@startlink[1]{}%
\providecommand \@@endlink[0]{}%
\providecommand \url  [0]{\begingroup\@sanitize@url \@url }%
\providecommand \@url [1]{\endgroup\@href {#1}{\urlprefix }}%
\providecommand \urlprefix  [0]{URL }%
\providecommand \Eprint [0]{\href }%
\providecommand \doibase [0]{http://dx.doi.org/}%
\providecommand \selectlanguage [0]{\@gobble}%
\providecommand \bibinfo  [0]{\@secondoftwo}%
\providecommand \bibfield  [0]{\@secondoftwo}%
\providecommand \translation [1]{[#1]}%
\providecommand \BibitemOpen [0]{}%
\providecommand \bibitemStop [0]{}%
\providecommand \bibitemNoStop [0]{.\EOS\space}%
\providecommand \EOS [0]{\spacefactor3000\relax}%
\providecommand \BibitemShut  [1]{\csname bibitem#1\endcsname}%
\let\auto@bib@innerbib\@empty
\bibitem [{\citenamefont {Bloch}\ \emph {et~al.}(2008)\citenamefont {Bloch},
  \citenamefont {Dalibard},\ and\ \citenamefont {Zwerger}}]{Bloch2008}%
  \BibitemOpen
  \bibfield  {author} {\bibinfo {author} {\bibfnamefont {Immanuel}\
  \bibnamefont {Bloch}}, \bibinfo {author} {\bibfnamefont {Jean}\ \bibnamefont
  {Dalibard}}, \ and\ \bibinfo {author} {\bibfnamefont {Wilhelm}\ \bibnamefont
  {Zwerger}},\ }\bibfield  {title} {\enquote {\bibinfo {title} {Many-body
  physics with ultracold gases},}\ }\href {\doibase 10.1103/RevModPhys.80.885}
  {\bibfield  {journal} {\bibinfo  {journal} {Rev. Mod. Phys.}\ }\textbf
  {\bibinfo {volume} {80}},\ \bibinfo {pages} {885--964} (\bibinfo {year}
  {2008})}\BibitemShut {NoStop}%
\bibitem [{\citenamefont {Bloch}\ \emph {et~al.}(2012)\citenamefont {Bloch},
  \citenamefont {Dalibard},\ and\ \citenamefont {Nascimbene}}]{Bloch2012}%
  \BibitemOpen
  \bibfield  {author} {\bibinfo {author} {\bibfnamefont {Immanuel}\
  \bibnamefont {Bloch}}, \bibinfo {author} {\bibfnamefont {Jean}\ \bibnamefont
  {Dalibard}}, \ and\ \bibinfo {author} {\bibfnamefont {Sylvain}\ \bibnamefont
  {Nascimbene}},\ }\bibfield  {title} {\enquote {\bibinfo {title} {Quantum
  simulations with ultracold quantum gases},}\ }\href {\doibase
  10.1038/nphys2259} {\bibfield  {journal} {\bibinfo  {journal} {Nature
  Physics}\ }\textbf {\bibinfo {volume} {8}},\ \bibinfo {pages} {267--276}
  (\bibinfo {year} {2012})}\BibitemShut {NoStop}%
\bibitem [{\citenamefont {Yarmchuk}\ \emph {et~al.}(1979)\citenamefont
  {Yarmchuk}, \citenamefont {Gordon},\ and\ \citenamefont
  {Packard}}]{Yarmchuk1979}%
  \BibitemOpen
  \bibfield  {author} {\bibinfo {author} {\bibfnamefont {E.~J.}\ \bibnamefont
  {Yarmchuk}}, \bibinfo {author} {\bibfnamefont {M.~J.~V.}\ \bibnamefont
  {Gordon}}, \ and\ \bibinfo {author} {\bibfnamefont {R.~E.}\ \bibnamefont
  {Packard}},\ }\bibfield  {title} {\enquote {\bibinfo {title} {Observation of
  stationary vortex arrays in rotating superfluid helium},}\ }\href {\doibase
  10.1103/PhysRevLett.43.214} {\bibfield  {journal} {\bibinfo  {journal} {Phys.
  Rev. Lett.}\ }\textbf {\bibinfo {volume} {43}},\ \bibinfo {pages} {214--217}
  (\bibinfo {year} {1979})}\BibitemShut {NoStop}%
\bibitem [{\citenamefont {Meyer}\ and\ \citenamefont {Reif}(1958)}]{Meyer1958}%
  \BibitemOpen
  \bibfield  {author} {\bibinfo {author} {\bibfnamefont {Lothar}\ \bibnamefont
  {Meyer}}\ and\ \bibinfo {author} {\bibfnamefont {F.}~\bibnamefont {Reif}},\
  }\bibfield  {title} {\enquote {\bibinfo {title} {Mobilities of he ions in
  liquid helium},}\ }\href {\doibase 10.1103/PhysRev.110.279} {\bibfield
  {journal} {\bibinfo  {journal} {Phys. Rev.}\ }\textbf {\bibinfo {volume}
  {110}},\ \bibinfo {pages} {279--280} (\bibinfo {year} {1958})}\BibitemShut
  {NoStop}%
\bibitem [{\citenamefont {Atkins}(1959)}]{Atkins1959}%
  \BibitemOpen
  \bibfield  {author} {\bibinfo {author} {\bibfnamefont {K.~R.}\ \bibnamefont
  {Atkins}},\ }\bibfield  {title} {\enquote {\bibinfo {title} {Ions in liquid
  helium},}\ }\href {\doibase 10.1103/PhysRev.116.1339} {\bibfield  {journal}
  {\bibinfo  {journal} {Phys. Rev.}\ }\textbf {\bibinfo {volume} {116}},\
  \bibinfo {pages} {1339--1343} (\bibinfo {year} {1959})}\BibitemShut {NoStop}%
\bibitem [{\citenamefont {Gross}(1962)}]{Gross1962}%
  \BibitemOpen
  \bibfield  {author} {\bibinfo {author} {\bibfnamefont {E.P}\ \bibnamefont
  {Gross}},\ }\bibfield  {title} {\enquote {\bibinfo {title} {Motion of foreign
  bodies in boson systems},}\ }\href {\doibase
  https://doi.org/10.1016/0003-4916(62)90217-8} {\bibfield  {journal} {\bibinfo
   {journal} {Annals of Physics}\ }\textbf {\bibinfo {volume} {19}},\ \bibinfo
  {pages} {234 -- 253} (\bibinfo {year} {1962})}\BibitemShut {NoStop}%
\bibitem [{\citenamefont {Ahonen}\ \emph {et~al.}(1976)\citenamefont {Ahonen},
  \citenamefont {Kokko}, \citenamefont {Lounasmaa}, \citenamefont {Paalanen},
  \citenamefont {Richardson}, \citenamefont {Schoepe},\ and\ \citenamefont
  {Takano}}]{Ahonen1976}%
  \BibitemOpen
  \bibfield  {author} {\bibinfo {author} {\bibfnamefont {A.~I.}\ \bibnamefont
  {Ahonen}}, \bibinfo {author} {\bibfnamefont {J.}~\bibnamefont {Kokko}},
  \bibinfo {author} {\bibfnamefont {O.~V.}\ \bibnamefont {Lounasmaa}}, \bibinfo
  {author} {\bibfnamefont {M.~A.}\ \bibnamefont {Paalanen}}, \bibinfo {author}
  {\bibfnamefont {R.~C.}\ \bibnamefont {Richardson}}, \bibinfo {author}
  {\bibfnamefont {W.}~\bibnamefont {Schoepe}}, \ and\ \bibinfo {author}
  {\bibfnamefont {Y.}~\bibnamefont {Takano}},\ }\bibfield  {title} {\enquote
  {\bibinfo {title} {Mobility of negative ions in superfluid
  $^{3}\mathrm{He}$},}\ }\href {\doibase 10.1103/PhysRevLett.37.511} {\bibfield
   {journal} {\bibinfo  {journal} {Phys. Rev. Lett.}\ }\textbf {\bibinfo
  {volume} {37}},\ \bibinfo {pages} {511--515} (\bibinfo {year}
  {1976})}\BibitemShut {NoStop}%
\bibitem [{\citenamefont {Roach}\ \emph {et~al.}(1977)\citenamefont {Roach},
  \citenamefont {Ketterson},\ and\ \citenamefont {Roach}}]{Roach1977}%
  \BibitemOpen
  \bibfield  {author} {\bibinfo {author} {\bibfnamefont {Paul~D.}\ \bibnamefont
  {Roach}}, \bibinfo {author} {\bibfnamefont {J.~B.}\ \bibnamefont
  {Ketterson}}, \ and\ \bibinfo {author} {\bibfnamefont {Pat~R.}\ \bibnamefont
  {Roach}},\ }\bibfield  {title} {\enquote {\bibinfo {title} {Mobility of
  positive and negative ions in superfluid $^{3}\mathrm{He}$},}\ }\href
  {\doibase 10.1103/PhysRevLett.39.626} {\bibfield  {journal} {\bibinfo
  {journal} {Phys. Rev. Lett.}\ }\textbf {\bibinfo {volume} {39}},\ \bibinfo
  {pages} {626--629} (\bibinfo {year} {1977})}\BibitemShut {NoStop}%
\bibitem [{\citenamefont {Ahonen}\ \emph {et~al.}(1978)\citenamefont {Ahonen},
  \citenamefont {Kokko}, \citenamefont {Paalanen}, \citenamefont {Richardson},
  \citenamefont {Schoepe},\ and\ \citenamefont {Takano}}]{Ahonen1978}%
  \BibitemOpen
  \bibfield  {author} {\bibinfo {author} {\bibfnamefont {A.~I.}\ \bibnamefont
  {Ahonen}}, \bibinfo {author} {\bibfnamefont {J.}~\bibnamefont {Kokko}},
  \bibinfo {author} {\bibfnamefont {M.~A.}\ \bibnamefont {Paalanen}}, \bibinfo
  {author} {\bibfnamefont {R.~C.}\ \bibnamefont {Richardson}}, \bibinfo
  {author} {\bibfnamefont {W.}~\bibnamefont {Schoepe}}, \ and\ \bibinfo
  {author} {\bibfnamefont {Y.}~\bibnamefont {Takano}},\ }\bibfield  {title}
  {\enquote {\bibinfo {title} {Negative ion motion in normal and superfluid
  3he},}\ }\href {\doibase 10.1007/BF00115525} {\bibfield  {journal} {\bibinfo
  {journal} {Journal of Low Temperature Physics}\ }\textbf {\bibinfo {volume}
  {30}},\ \bibinfo {pages} {205--228} (\bibinfo {year} {1978})}\BibitemShut
  {NoStop}%
\bibitem [{\citenamefont {Salomaa}\ \emph {et~al.}(1980)\citenamefont
  {Salomaa}, \citenamefont {Pethick},\ and\ \citenamefont
  {Baym}}]{Salomaa1980}%
  \BibitemOpen
  \bibfield  {author} {\bibinfo {author} {\bibfnamefont {M.}~\bibnamefont
  {Salomaa}}, \bibinfo {author} {\bibfnamefont {C.~J.}\ \bibnamefont
  {Pethick}}, \ and\ \bibinfo {author} {\bibfnamefont {Gordon}\ \bibnamefont
  {Baym}},\ }\bibfield  {title} {\enquote {\bibinfo {title} {Mobility tensor of
  the electron bubble in superfluid $^{3}\mathrm{He}$-$a$},}\ }\href {\doibase
  10.1103/PhysRevLett.44.998} {\bibfield  {journal} {\bibinfo  {journal} {Phys.
  Rev. Lett.}\ }\textbf {\bibinfo {volume} {44}},\ \bibinfo {pages} {998--1001}
  (\bibinfo {year} {1980})}\BibitemShut {NoStop}%
\bibitem [{\citenamefont {Baym}\ \emph {et~al.}(1979)\citenamefont {Baym},
  \citenamefont {Pethick},\ and\ \citenamefont {Salomaa}}]{Baym1979}%
  \BibitemOpen
  \bibfield  {author} {\bibinfo {author} {\bibfnamefont {Gordon}\ \bibnamefont
  {Baym}}, \bibinfo {author} {\bibfnamefont {C.~J.}\ \bibnamefont {Pethick}}, \
  and\ \bibinfo {author} {\bibfnamefont {M.}~\bibnamefont {Salomaa}},\
  }\bibfield  {title} {\enquote {\bibinfo {title} {Mobility of negative ions in
  superfluid 3he-b},}\ }\href {\doibase 10.1007/BF00118715} {\bibfield
  {journal} {\bibinfo  {journal} {Journal of Low Temperature Physics}\ }\textbf
  {\bibinfo {volume} {36}},\ \bibinfo {pages} {431--466} (\bibinfo {year}
  {1979})}\BibitemShut {NoStop}%
\bibitem [{\citenamefont {Grier}\ \emph {et~al.}(2009)\citenamefont {Grier},
  \citenamefont {Cetina}, \citenamefont {Oru\ifmmode \check{c}\else
  \v{c}\fi{}evi\ifmmode~\acute{c}\else \'{c}\fi{}},\ and\ \citenamefont
  {Vuleti\ifmmode~\acute{c}\else \'{c}\fi{}}}]{Grier2009}%
  \BibitemOpen
  \bibfield  {author} {\bibinfo {author} {\bibfnamefont {Andrew~T.}\
  \bibnamefont {Grier}}, \bibinfo {author} {\bibfnamefont {Marko}\ \bibnamefont
  {Cetina}}, \bibinfo {author} {\bibfnamefont {Fedja}\ \bibnamefont
  {Oru\ifmmode \check{c}\else \v{c}\fi{}evi\ifmmode~\acute{c}\else
  \'{c}\fi{}}}, \ and\ \bibinfo {author} {\bibfnamefont {Vladan}\ \bibnamefont
  {Vuleti\ifmmode~\acute{c}\else \'{c}\fi{}}},\ }\bibfield  {title} {\enquote
  {\bibinfo {title} {Observation of cold collisions between trapped ions and
  trapped atoms},}\ }\href {\doibase 10.1103/PhysRevLett.102.223201} {\bibfield
   {journal} {\bibinfo  {journal} {Phys. Rev. Lett.}\ }\textbf {\bibinfo
  {volume} {102}},\ \bibinfo {pages} {223201} (\bibinfo {year}
  {2009})}\BibitemShut {NoStop}%
\bibitem [{\citenamefont {Zipkes}\ \emph {et~al.}(2010)\citenamefont {Zipkes},
  \citenamefont {Palzer}, \citenamefont {Sias},\ and\ \citenamefont
  {K{\"o}hl}}]{Zipkes2010}%
  \BibitemOpen
  \bibfield  {author} {\bibinfo {author} {\bibfnamefont {Christoph}\
  \bibnamefont {Zipkes}}, \bibinfo {author} {\bibfnamefont {Stefan}\
  \bibnamefont {Palzer}}, \bibinfo {author} {\bibfnamefont {Carlo}\
  \bibnamefont {Sias}}, \ and\ \bibinfo {author} {\bibfnamefont {Michael}\
  \bibnamefont {K{\"o}hl}},\ }\bibfield  {title} {\enquote {\bibinfo {title} {A
  trapped single ion inside a bose--einstein condensate},}\ }\href {\doibase
  10.1038/nature08865} {\bibfield  {journal} {\bibinfo  {journal} {Nature}\
  }\textbf {\bibinfo {volume} {464}},\ \bibinfo {pages} {388--391} (\bibinfo
  {year} {2010})}\BibitemShut {NoStop}%
\bibitem [{\citenamefont {H\"arter}\ \emph
  {et~al.}(2012{\natexlab{a}})\citenamefont {H\"arter}, \citenamefont
  {Kr\"ukow}, \citenamefont {Brunner}, \citenamefont {Schnitzler},
  \citenamefont {Schmid},\ and\ \citenamefont {Denschlag}}]{Harter2012}%
  \BibitemOpen
  \bibfield  {author} {\bibinfo {author} {\bibfnamefont {Arne}\ \bibnamefont
  {H\"arter}}, \bibinfo {author} {\bibfnamefont {Artjom}\ \bibnamefont
  {Kr\"ukow}}, \bibinfo {author} {\bibfnamefont {Andreas}\ \bibnamefont
  {Brunner}}, \bibinfo {author} {\bibfnamefont {Wolfgang}\ \bibnamefont
  {Schnitzler}}, \bibinfo {author} {\bibfnamefont {Stefan}\ \bibnamefont
  {Schmid}}, \ and\ \bibinfo {author} {\bibfnamefont {Johannes~Hecker}\
  \bibnamefont {Denschlag}},\ }\bibfield  {title} {\enquote {\bibinfo {title}
  {Single ion as a three-body reaction center in an ultracold atomic gas},}\
  }\href {\doibase 10.1103/PhysRevLett.109.123201} {\bibfield  {journal}
  {\bibinfo  {journal} {Phys. Rev. Lett.}\ }\textbf {\bibinfo {volume} {109}},\
  \bibinfo {pages} {123201} (\bibinfo {year} {2012}{\natexlab{a}})}\BibitemShut
  {NoStop}%
\bibitem [{\citenamefont {Ratschbacher}\ \emph {et~al.}(2012)\citenamefont
  {Ratschbacher}, \citenamefont {Zipkes}, \citenamefont {Sias},\ and\
  \citenamefont {K{\"o}hl}}]{Ratschbacher2012}%
  \BibitemOpen
  \bibfield  {author} {\bibinfo {author} {\bibfnamefont {Lothar}\ \bibnamefont
  {Ratschbacher}}, \bibinfo {author} {\bibfnamefont {Christoph}\ \bibnamefont
  {Zipkes}}, \bibinfo {author} {\bibfnamefont {Carlo}\ \bibnamefont {Sias}}, \
  and\ \bibinfo {author} {\bibfnamefont {Michael}\ \bibnamefont {K{\"o}hl}},\
  }\bibfield  {title} {\enquote {\bibinfo {title} {Controlling chemical
  reactions of a single particle},}\ }\href {\doibase 10.1038/nphys2373}
  {\bibfield  {journal} {\bibinfo  {journal} {Nature Physics}\ }\textbf
  {\bibinfo {volume} {8}},\ \bibinfo {pages} {649--652} (\bibinfo {year}
  {2012})}\BibitemShut {NoStop}%
\bibitem [{\citenamefont {Kleinbach}\ \emph {et~al.}(2018)\citenamefont
  {Kleinbach}, \citenamefont {Engel}, \citenamefont {Dieterle}, \citenamefont
  {L\"ow}, \citenamefont {Pfau},\ and\ \citenamefont
  {Meinert}}]{Kleinbach2018}%
  \BibitemOpen
  \bibfield  {author} {\bibinfo {author} {\bibfnamefont {K.~S.}\ \bibnamefont
  {Kleinbach}}, \bibinfo {author} {\bibfnamefont {F.}~\bibnamefont {Engel}},
  \bibinfo {author} {\bibfnamefont {T.}~\bibnamefont {Dieterle}}, \bibinfo
  {author} {\bibfnamefont {R.}~\bibnamefont {L\"ow}}, \bibinfo {author}
  {\bibfnamefont {T.}~\bibnamefont {Pfau}}, \ and\ \bibinfo {author}
  {\bibfnamefont {F.}~\bibnamefont {Meinert}},\ }\bibfield  {title} {\enquote
  {\bibinfo {title} {Ionic impurity in a bose-einstein condensate at
  submicrokelvin temperatures},}\ }\href {\doibase
  10.1103/PhysRevLett.120.193401} {\bibfield  {journal} {\bibinfo  {journal}
  {Phys. Rev. Lett.}\ }\textbf {\bibinfo {volume} {120}},\ \bibinfo {pages}
  {193401} (\bibinfo {year} {2018})}\BibitemShut {NoStop}%
\bibitem [{\citenamefont {Sikorsky}\ \emph {et~al.}(2018)\citenamefont
  {Sikorsky}, \citenamefont {Meir}, \citenamefont {Ben-shlomi}, \citenamefont
  {Akerman},\ and\ \citenamefont {Ozeri}}]{Sikorsky2018}%
  \BibitemOpen
  \bibfield  {author} {\bibinfo {author} {\bibfnamefont {Tomas}\ \bibnamefont
  {Sikorsky}}, \bibinfo {author} {\bibfnamefont {Ziv}\ \bibnamefont {Meir}},
  \bibinfo {author} {\bibfnamefont {Ruti}\ \bibnamefont {Ben-shlomi}}, \bibinfo
  {author} {\bibfnamefont {Nitzan}\ \bibnamefont {Akerman}}, \ and\ \bibinfo
  {author} {\bibfnamefont {Roee}\ \bibnamefont {Ozeri}},\ }\bibfield  {title}
  {\enquote {\bibinfo {title} {Spin-controlled atom--ion chemistry},}\ }\href
  {\doibase 10.1038/s41467-018-03373-y} {\bibfield  {journal} {\bibinfo
  {journal} {Nature Communications}\ }\textbf {\bibinfo {volume} {9}},\
  \bibinfo {pages} {920} (\bibinfo {year} {2018})}\BibitemShut {NoStop}%
\bibitem [{\citenamefont {Feldker}\ \emph {et~al.}(2020)\citenamefont
  {Feldker}, \citenamefont {F{\"u}rst}, \citenamefont {Hirzler}, \citenamefont
  {Ewald}, \citenamefont {Mazzanti}, \citenamefont {Wiater}, \citenamefont
  {Tomza},\ and\ \citenamefont {Gerritsma}}]{Feldker2020}%
  \BibitemOpen
  \bibfield  {author} {\bibinfo {author} {\bibfnamefont {T.}~\bibnamefont
  {Feldker}}, \bibinfo {author} {\bibfnamefont {H.}~\bibnamefont {F{\"u}rst}},
  \bibinfo {author} {\bibfnamefont {H.}~\bibnamefont {Hirzler}}, \bibinfo
  {author} {\bibfnamefont {N.~V.}\ \bibnamefont {Ewald}}, \bibinfo {author}
  {\bibfnamefont {M.}~\bibnamefont {Mazzanti}}, \bibinfo {author}
  {\bibfnamefont {D.}~\bibnamefont {Wiater}}, \bibinfo {author} {\bibfnamefont
  {M.}~\bibnamefont {Tomza}}, \ and\ \bibinfo {author} {\bibfnamefont
  {R.}~\bibnamefont {Gerritsma}},\ }\bibfield  {title} {\enquote {\bibinfo
  {title} {Buffer gas cooling of a trapped ion to the quantum regime},}\ }\href
  {\doibase 10.1038/s41567-019-0772-5} {\bibfield  {journal} {\bibinfo
  {journal} {Nature Physics}\ }\textbf {\bibinfo {volume} {16}},\ \bibinfo
  {pages} {413--416} (\bibinfo {year} {2020})}\BibitemShut {NoStop}%
\bibitem [{\citenamefont {Schmidt}\ \emph {et~al.}(2020)\citenamefont
  {Schmidt}, \citenamefont {Weckesser}, \citenamefont {Thielemann},
  \citenamefont {Schaetz},\ and\ \citenamefont {Karpa}}]{Schmidt2020}%
  \BibitemOpen
  \bibfield  {author} {\bibinfo {author} {\bibfnamefont {J.}~\bibnamefont
  {Schmidt}}, \bibinfo {author} {\bibfnamefont {P.}~\bibnamefont {Weckesser}},
  \bibinfo {author} {\bibfnamefont {F.}~\bibnamefont {Thielemann}}, \bibinfo
  {author} {\bibfnamefont {T.}~\bibnamefont {Schaetz}}, \ and\ \bibinfo
  {author} {\bibfnamefont {L.}~\bibnamefont {Karpa}},\ }\bibfield  {title}
  {\enquote {\bibinfo {title} {Optical traps for sympathetic cooling of ions
  with ultracold neutral atoms},}\ }\href {\doibase
  10.1103/PhysRevLett.124.053402} {\bibfield  {journal} {\bibinfo  {journal}
  {Phys. Rev. Lett.}\ }\textbf {\bibinfo {volume} {124}},\ \bibinfo {pages}
  {053402} (\bibinfo {year} {2020})}\BibitemShut {NoStop}%
\bibitem [{\citenamefont {Dieterle}\ \emph
  {et~al.}(2020{\natexlab{a}})\citenamefont {Dieterle}, \citenamefont
  {Berngruber}, \citenamefont {H{\"o}lzl}, \citenamefont {L{\"o}w},
  \citenamefont {Jachymski}, \citenamefont {Pfau},\ and\ \citenamefont
  {Meinert}}]{dieterle2020}%
  \BibitemOpen
  \bibfield  {author} {\bibinfo {author} {\bibfnamefont {Thomas}\ \bibnamefont
  {Dieterle}}, \bibinfo {author} {\bibfnamefont {Moritz}\ \bibnamefont
  {Berngruber}}, \bibinfo {author} {\bibfnamefont {Christian}\ \bibnamefont
  {H{\"o}lzl}}, \bibinfo {author} {\bibfnamefont {Robert}\ \bibnamefont
  {L{\"o}w}}, \bibinfo {author} {\bibfnamefont {Krzysztof}\ \bibnamefont
  {Jachymski}}, \bibinfo {author} {\bibfnamefont {Tilman}\ \bibnamefont
  {Pfau}}, \ and\ \bibinfo {author} {\bibfnamefont {Florian}\ \bibnamefont
  {Meinert}},\ }\href@noop {} {\enquote {\bibinfo {title} {Transport of a
  single cold ion immersed in a bose-einstein condensate},}\ } (\bibinfo {year}
  {2020}{\natexlab{a}}),\ \Eprint {http://arxiv.org/abs/2007.00309}
  {arXiv:2007.00309 [physics.atom-ph]} \BibitemShut {NoStop}%
\bibitem [{\citenamefont {Dieterle}\ \emph
  {et~al.}(2020{\natexlab{b}})\citenamefont {Dieterle}, \citenamefont
  {Berngruber}, \citenamefont {H\"olzl}, \citenamefont {L\"ow}, \citenamefont
  {Jachymski}, \citenamefont {Pfau},\ and\ \citenamefont
  {Meinert}}]{Dieterle2020b}%
  \BibitemOpen
  \bibfield  {author} {\bibinfo {author} {\bibfnamefont {T.}~\bibnamefont
  {Dieterle}}, \bibinfo {author} {\bibfnamefont {M.}~\bibnamefont
  {Berngruber}}, \bibinfo {author} {\bibfnamefont {C.}~\bibnamefont {H\"olzl}},
  \bibinfo {author} {\bibfnamefont {R.}~\bibnamefont {L\"ow}}, \bibinfo
  {author} {\bibfnamefont {K.}~\bibnamefont {Jachymski}}, \bibinfo {author}
  {\bibfnamefont {T.}~\bibnamefont {Pfau}}, \ and\ \bibinfo {author}
  {\bibfnamefont {F.}~\bibnamefont {Meinert}},\ }\bibfield  {title} {\enquote
  {\bibinfo {title} {Inelastic collision dynamics of a single cold ion immersed
  in a bose-einstein condensate},}\ }\href {\doibase
  10.1103/PhysRevA.102.041301} {\bibfield  {journal} {\bibinfo  {journal}
  {Phys. Rev. A}\ }\textbf {\bibinfo {volume} {102}},\ \bibinfo {pages}
  {041301} (\bibinfo {year} {2020}{\natexlab{b}})}\BibitemShut {NoStop}%
\bibitem [{\citenamefont {Casteels}\ \emph {et~al.}(2011)\citenamefont
  {Casteels}, \citenamefont {Tempere},\ and\ \citenamefont
  {Devreese}}]{Casteels2011}%
  \BibitemOpen
  \bibfield  {author} {\bibinfo {author} {\bibfnamefont {W.}~\bibnamefont
  {Casteels}}, \bibinfo {author} {\bibfnamefont {J.}~\bibnamefont {Tempere}}, \
  and\ \bibinfo {author} {\bibfnamefont {J.~T.}\ \bibnamefont {Devreese}},\
  }\bibfield  {title} {\enquote {\bibinfo {title} {Polaronic properties of an
  ion in a bose-einstein condensate in the strong-coupling limit},}\ }\href
  {\doibase 10.1007/s10909-010-0286-0} {\bibfield  {journal} {\bibinfo
  {journal} {Journal of Low Temperature Physics}\ }\textbf {\bibinfo {volume}
  {162}},\ \bibinfo {pages} {266--273} (\bibinfo {year} {2011})}\BibitemShut
  {NoStop}%
\bibitem [{\citenamefont {Gao}(2010)}]{Gao2010}%
  \BibitemOpen
  \bibfield  {author} {\bibinfo {author} {\bibfnamefont {Bo}~\bibnamefont
  {Gao}},\ }\bibfield  {title} {\enquote {\bibinfo {title} {Universal
  properties in ultracold ion-atom interactions},}\ }\href {\doibase
  10.1103/PhysRevLett.104.213201} {\bibfield  {journal} {\bibinfo  {journal}
  {Phys. Rev. Lett.}\ }\textbf {\bibinfo {volume} {104}},\ \bibinfo {pages}
  {213201} (\bibinfo {year} {2010})}\BibitemShut {NoStop}%
\bibitem [{\citenamefont {Kr\"ukow}\ \emph {et~al.}(2016)\citenamefont
  {Kr\"ukow}, \citenamefont {Mohammadi}, \citenamefont {H\"arter},
  \citenamefont {Denschlag}, \citenamefont {P\'erez-R\'{\i}os},\ and\
  \citenamefont {Greene}}]{Krukow2016}%
  \BibitemOpen
  \bibfield  {author} {\bibinfo {author} {\bibfnamefont {Artjom}\ \bibnamefont
  {Kr\"ukow}}, \bibinfo {author} {\bibfnamefont {Amir}\ \bibnamefont
  {Mohammadi}}, \bibinfo {author} {\bibfnamefont {Arne}\ \bibnamefont
  {H\"arter}}, \bibinfo {author} {\bibfnamefont {Johannes~Hecker}\ \bibnamefont
  {Denschlag}}, \bibinfo {author} {\bibfnamefont {Jes\'us}\ \bibnamefont
  {P\'erez-R\'{\i}os}}, \ and\ \bibinfo {author} {\bibfnamefont {Chris~H.}\
  \bibnamefont {Greene}},\ }\bibfield  {title} {\enquote {\bibinfo {title}
  {Energy scaling of cold atom-atom-ion three-body recombination},}\ }\href
  {\doibase 10.1103/PhysRevLett.116.193201} {\bibfield  {journal} {\bibinfo
  {journal} {Phys. Rev. Lett.}\ }\textbf {\bibinfo {volume} {116}},\ \bibinfo
  {pages} {193201} (\bibinfo {year} {2016})}\BibitemShut {NoStop}%
\bibitem [{\citenamefont {C\^ot\'e}\ \emph {et~al.}(2002)\citenamefont
  {C\^ot\'e}, \citenamefont {Kharchenko},\ and\ \citenamefont
  {Lukin}}]{Lukin2002}%
  \BibitemOpen
  \bibfield  {author} {\bibinfo {author} {\bibfnamefont {R.}~\bibnamefont
  {C\^ot\'e}}, \bibinfo {author} {\bibfnamefont {V.}~\bibnamefont
  {Kharchenko}}, \ and\ \bibinfo {author} {\bibfnamefont {M.~D.}\ \bibnamefont
  {Lukin}},\ }\bibfield  {title} {\enquote {\bibinfo {title} {Mesoscopic
  molecular ions in bose-einstein condensates},}\ }\href {\doibase
  10.1103/PhysRevLett.89.093001} {\bibfield  {journal} {\bibinfo  {journal}
  {Phys. Rev. Lett.}\ }\textbf {\bibinfo {volume} {89}},\ \bibinfo {pages}
  {093001} (\bibinfo {year} {2002})}\BibitemShut {NoStop}%
\bibitem [{\citenamefont {Massignan}\ \emph {et~al.}(2005)\citenamefont
  {Massignan}, \citenamefont {Pethick},\ and\ \citenamefont
  {Smith}}]{Massignan2005}%
  \BibitemOpen
  \bibfield  {author} {\bibinfo {author} {\bibfnamefont {P.}~\bibnamefont
  {Massignan}}, \bibinfo {author} {\bibfnamefont {C.~J.}\ \bibnamefont
  {Pethick}}, \ and\ \bibinfo {author} {\bibfnamefont {H.}~\bibnamefont
  {Smith}},\ }\bibfield  {title} {\enquote {\bibinfo {title} {Static properties
  of positive ions in atomic bose-einstein condensates},}\ }\href {\doibase
  10.1103/PhysRevA.71.023606} {\bibfield  {journal} {\bibinfo  {journal} {Phys.
  Rev. A}\ }\textbf {\bibinfo {volume} {71}},\ \bibinfo {pages} {023606}
  (\bibinfo {year} {2005})}\BibitemShut {NoStop}%
\bibitem [{\citenamefont {Schurer}\ \emph {et~al.}(2017)\citenamefont
  {Schurer}, \citenamefont {Negretti},\ and\ \citenamefont
  {Schmelcher}}]{Schurer2017}%
  \BibitemOpen
  \bibfield  {author} {\bibinfo {author} {\bibfnamefont {J.~M.}\ \bibnamefont
  {Schurer}}, \bibinfo {author} {\bibfnamefont {A.}~\bibnamefont {Negretti}}, \
  and\ \bibinfo {author} {\bibfnamefont {P.}~\bibnamefont {Schmelcher}},\
  }\bibfield  {title} {\enquote {\bibinfo {title} {Unraveling the structure of
  ultracold mesoscopic collinear molecular ions},}\ }\href {\doibase
  10.1103/PhysRevLett.119.063001} {\bibfield  {journal} {\bibinfo  {journal}
  {Phys. Rev. Lett.}\ }\textbf {\bibinfo {volume} {119}},\ \bibinfo {pages}
  {063001} (\bibinfo {year} {2017})}\BibitemShut {NoStop}%
\bibitem [{\citenamefont {Astrakharchik}\ \emph {et~al.}(2020)\citenamefont
  {Astrakharchik}, \citenamefont {Ardila}, \citenamefont {Schmidt},
  \citenamefont {Jachymski},\ and\ \citenamefont
  {Negretti}}]{astrakharchik2020}%
  \BibitemOpen
  \bibfield  {author} {\bibinfo {author} {\bibfnamefont {G.~E.}\ \bibnamefont
  {Astrakharchik}}, \bibinfo {author} {\bibfnamefont {L.~A.~Pe{\~n}a}\
  \bibnamefont {Ardila}}, \bibinfo {author} {\bibfnamefont {R.}~\bibnamefont
  {Schmidt}}, \bibinfo {author} {\bibfnamefont {K.}~\bibnamefont {Jachymski}},
  \ and\ \bibinfo {author} {\bibfnamefont {A.}~\bibnamefont {Negretti}},\
  }\href@noop {} {\enquote {\bibinfo {title} {Ionic polaron in a bose-einstein
  condensate},}\ } (\bibinfo {year} {2020}),\ \Eprint
  {http://arxiv.org/abs/2005.12033} {arXiv:2005.12033 [cond-mat.quant-gas]}
  \BibitemShut {NoStop}%
\bibitem [{\citenamefont {Tomza}\ \emph {et~al.}(2019)\citenamefont {Tomza},
  \citenamefont {Jachymski}, \citenamefont {Gerritsma}, \citenamefont
  {Negretti}, \citenamefont {Calarco}, \citenamefont {Idziaszek},\ and\
  \citenamefont {Julienne}}]{Tomza2019}%
  \BibitemOpen
  \bibfield  {author} {\bibinfo {author} {\bibfnamefont {Micha\l{}}\
  \bibnamefont {Tomza}}, \bibinfo {author} {\bibfnamefont {Krzysztof}\
  \bibnamefont {Jachymski}}, \bibinfo {author} {\bibfnamefont {Rene}\
  \bibnamefont {Gerritsma}}, \bibinfo {author} {\bibfnamefont {Antonio}\
  \bibnamefont {Negretti}}, \bibinfo {author} {\bibfnamefont {Tommaso}\
  \bibnamefont {Calarco}}, \bibinfo {author} {\bibfnamefont {Zbigniew}\
  \bibnamefont {Idziaszek}}, \ and\ \bibinfo {author} {\bibfnamefont {Paul~S.}\
  \bibnamefont {Julienne}},\ }\bibfield  {title} {\enquote {\bibinfo {title}
  {Cold hybrid ion-atom systems},}\ }\href {\doibase
  10.1103/RevModPhys.91.035001} {\bibfield  {journal} {\bibinfo  {journal}
  {Rev. Mod. Phys.}\ }\textbf {\bibinfo {volume} {91}},\ \bibinfo {pages}
  {035001} (\bibinfo {year} {2019})}\BibitemShut {NoStop}%
\bibitem [{\citenamefont {Krych}\ and\ \citenamefont
  {Idziaszek}(2015)}]{Krych2015}%
  \BibitemOpen
  \bibfield  {author} {\bibinfo {author} {\bibfnamefont {Micha\l{}}\
  \bibnamefont {Krych}}\ and\ \bibinfo {author} {\bibfnamefont {Zbigniew}\
  \bibnamefont {Idziaszek}},\ }\bibfield  {title} {\enquote {\bibinfo {title}
  {Description of ion motion in a paul trap immersed in a cold atomic gas},}\
  }\href {\doibase 10.1103/PhysRevA.91.023430} {\bibfield  {journal} {\bibinfo
  {journal} {Phys. Rev. A}\ }\textbf {\bibinfo {volume} {91}},\ \bibinfo
  {pages} {023430} (\bibinfo {year} {2015})}\BibitemShut {NoStop}%
\bibitem [{\citenamefont {Lee}\ \emph {et~al.}(1953)\citenamefont {Lee},
  \citenamefont {Low},\ and\ \citenamefont {Pines}}]{Lee1953}%
  \BibitemOpen
  \bibfield  {author} {\bibinfo {author} {\bibfnamefont {T.~D.}\ \bibnamefont
  {Lee}}, \bibinfo {author} {\bibfnamefont {F.~E.}\ \bibnamefont {Low}}, \ and\
  \bibinfo {author} {\bibfnamefont {D.}~\bibnamefont {Pines}},\ }\bibfield
  {title} {\enquote {\bibinfo {title} {The motion of slow electrons in a polar
  crystal},}\ }\href {\doibase 10.1103/PhysRev.90.297} {\bibfield  {journal}
  {\bibinfo  {journal} {Phys. Rev.}\ }\textbf {\bibinfo {volume} {90}},\
  \bibinfo {pages} {297--302} (\bibinfo {year} {1953})}\BibitemShut {NoStop}%
\bibitem [{\citenamefont {Shashi}\ \emph {et~al.}(2014)\citenamefont {Shashi},
  \citenamefont {Grusdt}, \citenamefont {Abanin},\ and\ \citenamefont
  {Demler}}]{Shashi2014}%
  \BibitemOpen
  \bibfield  {author} {\bibinfo {author} {\bibfnamefont {Aditya}\ \bibnamefont
  {Shashi}}, \bibinfo {author} {\bibfnamefont {Fabian}\ \bibnamefont {Grusdt}},
  \bibinfo {author} {\bibfnamefont {Dmitry~A.}\ \bibnamefont {Abanin}}, \ and\
  \bibinfo {author} {\bibfnamefont {Eugene}\ \bibnamefont {Demler}},\
  }\bibfield  {title} {\enquote {\bibinfo {title} {Radio-frequency spectroscopy
  of polarons in ultracold bose gases},}\ }\href {\doibase
  10.1103/PhysRevA.89.053617} {\bibfield  {journal} {\bibinfo  {journal} {Phys.
  Rev. A}\ }\textbf {\bibinfo {volume} {89}},\ \bibinfo {pages} {053617}
  (\bibinfo {year} {2014})}\BibitemShut {NoStop}%
\bibitem [{\citenamefont {Shchadilova}\ \emph {et~al.}(2016)\citenamefont
  {Shchadilova}, \citenamefont {Schmidt}, \citenamefont {Grusdt},\ and\
  \citenamefont {Demler}}]{Shchadilova2016}%
  \BibitemOpen
  \bibfield  {author} {\bibinfo {author} {\bibfnamefont {Yulia~E.}\
  \bibnamefont {Shchadilova}}, \bibinfo {author} {\bibfnamefont {Richard}\
  \bibnamefont {Schmidt}}, \bibinfo {author} {\bibfnamefont {Fabian}\
  \bibnamefont {Grusdt}}, \ and\ \bibinfo {author} {\bibfnamefont {Eugene}\
  \bibnamefont {Demler}},\ }\bibfield  {title} {\enquote {\bibinfo {title}
  {Quantum dynamics of ultracold bose polarons},}\ }\href {\doibase
  10.1103/PhysRevLett.117.113002} {\bibfield  {journal} {\bibinfo  {journal}
  {Phys. Rev. Lett.}\ }\textbf {\bibinfo {volume} {117}},\ \bibinfo {pages}
  {113002} (\bibinfo {year} {2016})}\BibitemShut {NoStop}%
\bibitem [{SM()}]{SM}%
  \BibitemOpen
  \href@noop {} {}\bibinfo {note} {See {\it Supplemental Material} online for
  details.}\BibitemShut {Stop}%
\bibitem [{\citenamefont {Knap}\ \emph {et~al.}(2012)\citenamefont {Knap},
  \citenamefont {Shashi}, \citenamefont {Nishida}, \citenamefont {Imambekov},
  \citenamefont {Abanin},\ and\ \citenamefont {Demler}}]{Knap2012}%
  \BibitemOpen
  \bibfield  {author} {\bibinfo {author} {\bibfnamefont {Michael}\ \bibnamefont
  {Knap}}, \bibinfo {author} {\bibfnamefont {Aditya}\ \bibnamefont {Shashi}},
  \bibinfo {author} {\bibfnamefont {Yusuke}\ \bibnamefont {Nishida}}, \bibinfo
  {author} {\bibfnamefont {Adilet}\ \bibnamefont {Imambekov}}, \bibinfo
  {author} {\bibfnamefont {Dmitry~A.}\ \bibnamefont {Abanin}}, \ and\ \bibinfo
  {author} {\bibfnamefont {Eugene}\ \bibnamefont {Demler}},\ }\bibfield
  {title} {\enquote {\bibinfo {title} {Time-dependent impurity in ultracold
  fermions: Orthogonality catastrophe and beyond},}\ }\href {\doibase
  10.1103/PhysRevX.2.041020} {\bibfield  {journal} {\bibinfo  {journal} {Phys.
  Rev. X}\ }\textbf {\bibinfo {volume} {2}},\ \bibinfo {pages} {041020}
  (\bibinfo {year} {2012})}\BibitemShut {NoStop}%
\bibitem [{\citenamefont {Massignan}\ and\ \citenamefont
  {Bruun}(2011)}]{Massignan2011}%
  \BibitemOpen
  \bibfield  {author} {\bibinfo {author} {\bibfnamefont {P.}~\bibnamefont
  {Massignan}}\ and\ \bibinfo {author} {\bibfnamefont {G.M.}\ \bibnamefont
  {Bruun}},\ }\bibfield  {title} {\enquote {\bibinfo {title} {Repulsive
  polarons and itinerant ferromagnetism in strongly polarized fermi gases},}\
  }\href {\doibase 10.1140/epjd/e2011-20084-5} {\bibfield  {journal} {\bibinfo
  {journal} {EPJ D}\ }\textbf {\bibinfo {volume} {65}},\ \bibinfo {pages}
  {83--89} (\bibinfo {year} {2011})}\BibitemShut {NoStop}%
\bibitem [{\citenamefont {J\o{}rgensen}\ \emph {et~al.}(2016)\citenamefont
  {J\o{}rgensen}, \citenamefont {Wacker}, \citenamefont {Skalmstang},
  \citenamefont {Parish}, \citenamefont {Levinsen}, \citenamefont
  {Christensen}, \citenamefont {Bruun},\ and\ \citenamefont
  {Arlt}}]{Jorgensen2016}%
  \BibitemOpen
  \bibfield  {author} {\bibinfo {author} {\bibfnamefont {Nils~B.}\ \bibnamefont
  {J\o{}rgensen}}, \bibinfo {author} {\bibfnamefont {Lars}\ \bibnamefont
  {Wacker}}, \bibinfo {author} {\bibfnamefont {Kristoffer~T.}\ \bibnamefont
  {Skalmstang}}, \bibinfo {author} {\bibfnamefont {Meera~M.}\ \bibnamefont
  {Parish}}, \bibinfo {author} {\bibfnamefont {Jesper}\ \bibnamefont
  {Levinsen}}, \bibinfo {author} {\bibfnamefont {Rasmus~S.}\ \bibnamefont
  {Christensen}}, \bibinfo {author} {\bibfnamefont {Georg~M.}\ \bibnamefont
  {Bruun}}, \ and\ \bibinfo {author} {\bibfnamefont {Jan~J.}\ \bibnamefont
  {Arlt}},\ }\bibfield  {title} {\enquote {\bibinfo {title} {Observation of
  attractive and repulsive polarons in a bose-einstein condensate},}\ }\href
  {\doibase 10.1103/PhysRevLett.117.055302} {\bibfield  {journal} {\bibinfo
  {journal} {Phys. Rev. Lett.}\ }\textbf {\bibinfo {volume} {117}},\ \bibinfo
  {pages} {055302} (\bibinfo {year} {2016})}\BibitemShut {NoStop}%
\bibitem [{\citenamefont {Hu}\ \emph {et~al.}(2016)\citenamefont {Hu},
  \citenamefont {Van~de Graaff}, \citenamefont {Kedar}, \citenamefont {Corson},
  \citenamefont {Cornell},\ and\ \citenamefont {Jin}}]{Hu2016}%
  \BibitemOpen
  \bibfield  {author} {\bibinfo {author} {\bibfnamefont {Ming-Guang}\
  \bibnamefont {Hu}}, \bibinfo {author} {\bibfnamefont {Michael~J.}\
  \bibnamefont {Van~de Graaff}}, \bibinfo {author} {\bibfnamefont {Dhruv}\
  \bibnamefont {Kedar}}, \bibinfo {author} {\bibfnamefont {John~P.}\
  \bibnamefont {Corson}}, \bibinfo {author} {\bibfnamefont {Eric~A.}\
  \bibnamefont {Cornell}}, \ and\ \bibinfo {author} {\bibfnamefont
  {Deborah~S.}\ \bibnamefont {Jin}},\ }\bibfield  {title} {\enquote {\bibinfo
  {title} {Bose polarons in the strongly interacting regime},}\ }\href
  {\doibase 10.1103/PhysRevLett.117.055301} {\bibfield  {journal} {\bibinfo
  {journal} {Phys. Rev. Lett.}\ }\textbf {\bibinfo {volume} {117}},\ \bibinfo
  {pages} {055301} (\bibinfo {year} {2016})}\BibitemShut {NoStop}%
\bibitem [{\citenamefont {Pe\~na Ardila}\ \emph {et~al.}(2019)\citenamefont
  {Pe\~na Ardila}, \citenamefont {J\o{}rgensen}, \citenamefont {Pohl},
  \citenamefont {Giorgini}, \citenamefont {Bruun},\ and\ \citenamefont
  {Arlt}}]{Ardila2019}%
  \BibitemOpen
  \bibfield  {author} {\bibinfo {author} {\bibfnamefont {L.~A.}\ \bibnamefont
  {Pe\~na Ardila}}, \bibinfo {author} {\bibfnamefont {N.~B.}\ \bibnamefont
  {J\o{}rgensen}}, \bibinfo {author} {\bibfnamefont {T.}~\bibnamefont {Pohl}},
  \bibinfo {author} {\bibfnamefont {S.}~\bibnamefont {Giorgini}}, \bibinfo
  {author} {\bibfnamefont {G.~M.}\ \bibnamefont {Bruun}}, \ and\ \bibinfo
  {author} {\bibfnamefont {J.~J.}\ \bibnamefont {Arlt}},\ }\bibfield  {title}
  {\enquote {\bibinfo {title} {Analyzing a bose polaron across resonant
  interactions},}\ }\href {\doibase 10.1103/PhysRevA.99.063607} {\bibfield
  {journal} {\bibinfo  {journal} {Phys. Rev. A}\ }\textbf {\bibinfo {volume}
  {99}},\ \bibinfo {pages} {063607} (\bibinfo {year} {2019})}\BibitemShut
  {NoStop}%
\bibitem [{\citenamefont {Yan}\ \emph {et~al.}(2020)\citenamefont {Yan},
  \citenamefont {Ni}, \citenamefont {Robens},\ and\ \citenamefont
  {Zwierlein}}]{Yan2020}%
  \BibitemOpen
  \bibfield  {author} {\bibinfo {author} {\bibfnamefont {Zoe~Z.}\ \bibnamefont
  {Yan}}, \bibinfo {author} {\bibfnamefont {Yiqi}\ \bibnamefont {Ni}}, \bibinfo
  {author} {\bibfnamefont {Carsten}\ \bibnamefont {Robens}}, \ and\ \bibinfo
  {author} {\bibfnamefont {Martin~W.}\ \bibnamefont {Zwierlein}},\ }\bibfield
  {title} {\enquote {\bibinfo {title} {Bose polarons near quantum
  criticality},}\ }\href {\doibase 10.1126/science.aax5850} {\bibfield
  {journal} {\bibinfo  {journal} {Science}\ }\textbf {\bibinfo {volume}
  {368}},\ \bibinfo {pages} {190--194} (\bibinfo {year} {2020})},\ \Eprint
  {http://arxiv.org/abs/https://science.sciencemag.org/content/368/6487/190.full.pdf}
  {https://science.sciencemag.org/content/368/6487/190.full.pdf} \BibitemShut
  {NoStop}%
\bibitem [{foo()}]{footnote}%
  \BibitemOpen
  \href@noop {} {}\bibinfo {note} {In the top panel of
  Fig.~\ref{comparenobnd1}, we take the logarithm to the spectral function,
  which visualy makes the width appear larger.}\BibitemShut {Stop}%
\bibitem [{\citenamefont {Rath}\ and\ \citenamefont
  {Schmidt}(2013)}]{Rath2013}%
  \BibitemOpen
  \bibfield  {author} {\bibinfo {author} {\bibfnamefont {Steffen~Patrick}\
  \bibnamefont {Rath}}\ and\ \bibinfo {author} {\bibfnamefont {Richard}\
  \bibnamefont {Schmidt}},\ }\bibfield  {title} {\enquote {\bibinfo {title}
  {Field-theoretical study of the bose polaron},}\ }\href {\doibase
  10.1103/PhysRevA.88.053632} {\bibfield  {journal} {\bibinfo  {journal} {Phys.
  Rev. A}\ }\textbf {\bibinfo {volume} {88}},\ \bibinfo {pages} {053632}
  (\bibinfo {year} {2013})}\BibitemShut {NoStop}%
\bibitem [{Pie()}]{PietroPrivate}%
  \BibitemOpen
  \href@noop {} {}\bibinfo {note} {We thank P.\ Massignan for pointing this
  result out for us}\BibitemShut {NoStop}%
\bibitem [{\citenamefont {Nielsen}\ \emph {et~al.}(2019)\citenamefont
  {Nielsen}, \citenamefont {Ardila}, \citenamefont {Bruun},\ and\ \citenamefont
  {Pohl}}]{Nielsen2019}%
  \BibitemOpen
  \bibfield  {author} {\bibinfo {author} {\bibfnamefont {K~Knakkergaard}\
  \bibnamefont {Nielsen}}, \bibinfo {author} {\bibfnamefont {L~A~Pe{\~{n}}a}\
  \bibnamefont {Ardila}}, \bibinfo {author} {\bibfnamefont {G~M}\ \bibnamefont
  {Bruun}}, \ and\ \bibinfo {author} {\bibfnamefont {T}~\bibnamefont {Pohl}},\
  }\bibfield  {title} {\enquote {\bibinfo {title} {Critical slowdown of
  non-equilibrium polaron dynamics},}\ }\href {\doibase
  10.1088/1367-2630/ab0a81} {\bibfield  {journal} {\bibinfo  {journal} {New
  Journal of Physics}\ }\textbf {\bibinfo {volume} {21}},\ \bibinfo {pages}
  {043014} (\bibinfo {year} {2019})}\BibitemShut {NoStop}%
\bibitem [{\citenamefont {H\"arter}\ \emph
  {et~al.}(2012{\natexlab{b}})\citenamefont {H\"arter}, \citenamefont
  {Kr\"ukow}, \citenamefont {Brunner}, \citenamefont {Schnitzler},
  \citenamefont {Schmid},\ and\ \citenamefont {Denschlag}}]{Harter2021}%
  \BibitemOpen
  \bibfield  {author} {\bibinfo {author} {\bibfnamefont {Arne}\ \bibnamefont
  {H\"arter}}, \bibinfo {author} {\bibfnamefont {Artjom}\ \bibnamefont
  {Kr\"ukow}}, \bibinfo {author} {\bibfnamefont {Andreas}\ \bibnamefont
  {Brunner}}, \bibinfo {author} {\bibfnamefont {Wolfgang}\ \bibnamefont
  {Schnitzler}}, \bibinfo {author} {\bibfnamefont {Stefan}\ \bibnamefont
  {Schmid}}, \ and\ \bibinfo {author} {\bibfnamefont {Johannes~Hecker}\
  \bibnamefont {Denschlag}},\ }\bibfield  {title} {\enquote {\bibinfo {title}
  {Single ion as a three-body reaction center in an ultracold atomic gas},}\
  }\href {\doibase 10.1103/PhysRevLett.109.123201} {\bibfield  {journal}
  {\bibinfo  {journal} {Phys. Rev. Lett.}\ }\textbf {\bibinfo {volume} {109}},\
  \bibinfo {pages} {123201} (\bibinfo {year} {2012}{\natexlab{b}})}\BibitemShut
  {NoStop}%
\bibitem [{\citenamefont {{Skou}}\ \emph {et~al.}(2020)\citenamefont {{Skou}},
  \citenamefont {{Skov}}, \citenamefont {{J{\o}rgensen}}, \citenamefont
  {{Nielsen}}, \citenamefont {{Camacho-Guardian}}, \citenamefont {{Pohl}},
  \citenamefont {{Bruun}},\ and\ \citenamefont {{Arlt}}}]{Skou2020}%
  \BibitemOpen
  \bibfield  {author} {\bibinfo {author} {\bibfnamefont {Magnus~G.}\
  \bibnamefont {{Skou}}}, \bibinfo {author} {\bibfnamefont {Thomas~G.}\
  \bibnamefont {{Skov}}}, \bibinfo {author} {\bibfnamefont {Nils~B.}\
  \bibnamefont {{J{\o}rgensen}}}, \bibinfo {author} {\bibfnamefont
  {Kristian~K.}\ \bibnamefont {{Nielsen}}}, \bibinfo {author} {\bibfnamefont
  {Arturo}\ \bibnamefont {{Camacho-Guardian}}}, \bibinfo {author}
  {\bibfnamefont {Thomas}\ \bibnamefont {{Pohl}}}, \bibinfo {author}
  {\bibfnamefont {Georg~M.}\ \bibnamefont {{Bruun}}}, \ and\ \bibinfo {author}
  {\bibfnamefont {Jan~J.}\ \bibnamefont {{Arlt}}},\ }\bibfield  {title}
  {\enquote {\bibinfo {title} {{Non-equilibrium dynamics of quantum
  impurities}},}\ }\href@noop {} {\bibfield  {journal} {\bibinfo  {journal}
  {arXiv e-prints}\ ,\ \bibinfo {eid} {arXiv:2005.00424}} (\bibinfo {year}
  {2020})},\ \Eprint {http://arxiv.org/abs/2005.00424} {arXiv:2005.00424
  [cond-mat.quant-gas]} \BibitemShut {NoStop}%
\bibitem [{\citenamefont {Dzsotjan}\ \emph {et~al.}(2020)\citenamefont
  {Dzsotjan}, \citenamefont {Schmidt},\ and\ \citenamefont
  {Fleischhauer}}]{Dzsotjan2019}%
  \BibitemOpen
  \bibfield  {author} {\bibinfo {author} {\bibfnamefont {David}\ \bibnamefont
  {Dzsotjan}}, \bibinfo {author} {\bibfnamefont {Richard}\ \bibnamefont
  {Schmidt}}, \ and\ \bibinfo {author} {\bibfnamefont {Michael}\ \bibnamefont
  {Fleischhauer}},\ }\bibfield  {title} {\enquote {\bibinfo {title} {Dynamical
  variational approach to bose polarons at finite temperatures},}\ }\href
  {\doibase 10.1103/PhysRevLett.124.223401} {\bibfield  {journal} {\bibinfo
  {journal} {Phys. Rev. Lett.}\ }\textbf {\bibinfo {volume} {124}},\ \bibinfo
  {pages} {223401} (\bibinfo {year} {2020})}\BibitemShut {NoStop}%
\bibitem [{\citenamefont {Dishan}(1995)}]{Dishan1995}%
  \BibitemOpen
  \bibfield  {author} {\bibinfo {author} {\bibfnamefont {Huang}\ \bibnamefont
  {Dishan}},\ }\bibfield  {title} {\enquote {\bibinfo {title} {Phase error in
  fast fourier transform analysis},}\ }\href {\doibase
  https://doi.org/10.1006/mssp.1995.0009} {\bibfield  {journal} {\bibinfo
  {journal} {Mechanical Systems and Signal Processing}\ }\textbf {\bibinfo
  {volume} {9}},\ \bibinfo {pages} {113 -- 118} (\bibinfo {year}
  {1995})}\BibitemShut {NoStop}%
\end{thebibliography}%


\begin{widetext}
\section{Supplemental Material}
\subsection{Equations of motion}
The Hamiltonian in Eq. [1] (main text) written explicitly in terms of the position $\hat{\mathbf{r}}$ and momentum $\hat{\mathbf{P}}$ operators of the ion reads
 \begin{gather}
\hat{H} =  \frac{\hat{\mathbf{P}}^2}{2m} + \sum_{\mathbf{k}}\epsilon_{\mathbf{k}} \hat{b}^\dagger_\mathbf{k} \hat{b}_\mathbf{k} + \frac{g_{B}}{2} \sum_{\mathbf{k},\mathbf{k'},\mathbf{q}} \hat{b}^\dagger_\mathbf{k'+q} \hat{b}^\dagger_\mathbf{k-q}\hat{b}_\mathbf{k'} \hat{b}_\mathbf{k} + \sum_{\mathbf{k},\mathbf{q}} V(\mathbf{q}) e^{-i\mathbf{q}\cdot \hat{\mathbf{r}}} \hat{b}^\dagger_{\mathbf{k+q}} \hat{b}_\mathbf{k}.
\end{gather}
We perform the following Bogoliubov transformation 
\begin{align}
     & \hat{\beta}_\mathbf{k} = u_\mathbf{k} \hat b_\mathbf{k} - v_\mathbf{k} \hat b_{\mathbf{-k}}^\dagger, \ \hat{\beta}_\mathbf{k}^\dagger = u_\mathbf{k} \hat b_\mathbf{k}^\dagger - v_\mathbf{k} \hat b_{\mathbf{-k}},
 \end{align}
 where $\mathbf{k} \neq \mathbf{0},$ and under the standard Bogoliubov approximations we arrive at an extended Fr\"ohlich Hamiltonian that takes the form
\begin{gather}
        \hat{H} = \frac{\hat{\mathbf{P}}^2}{2m} + n_0 V(\mathbf{0}) + \sum_{\mathbf{k\neq \mathbf{0}}} \omega_{\mathbf{k}} \hat \beta^\dagger_{\mathbf{k}} \hat \beta_{\mathbf{k}}  + \sqrt{n_0} \sum_{\mathbf{q}} \sqrt{\epsilon_\mathbf{q}/\omega_\mathbf{q}} V(\mathbf{q}) e^{-i\mathbf{q}\cdot \hat{\mathbf{r}}}\left( \hat \beta_\mathbf{q}^\dagger + \hat \beta_{\mathbf{-q}} \right)\\ \nonumber
         +  \sum_{\mathbf{k}\mathbf{k'}}  V(\mathbf{k}-\mathbf{k}') e^{i (\mathbf{k'}-\mathbf{k})\cdot \hat{\mathbf{r}}} (u_{\mathbf{k'}} u_{\mathbf{k}} + v_{\mathbf{k'}} v_{\mathbf{k}}) \hat \beta^\dagger_\mathbf{k} \hat \beta_{\mathbf{k'}} \\ \nonumber
         - \sum_{\mathbf{k}\mathbf{k'}}  V(\mathbf{k}'+\mathbf{k}) e^{-i (\mathbf{k}+\mathbf{k}')\cdot \hat{\mathbf{r}}} (v_{\mathbf{k'}} u_\mathbf{k}+ u_\mathbf{k'}v_{\mathbf{k}}) (\hat \beta_{\mathbf{k}}^\dagger \hat \beta_\mathbf{k'}^\dagger + \hat \beta_{-\mathbf{k}} \hat \beta_\mathbf{-k'} ).
    \end{gather}
 It is convenient to employ the following canonical transformation
\begin{align}
    \hat{S} = e^{-i \hat{\mathbf{r}}\cdot \hat{\mathbf{P}}_B} = e^{-i\hat{\mathbf{r}}\cdot \sum_\mathbf{k} \mathbf{k} \beta^\dagger_\mathbf{k} \beta_\mathbf{k}},
\end{align}
since the transformation eliminates the coordinates of the ion and the operator $\hat{\mathbf{P}}$ commutes with the final form of the Hamiltonian:
\begin{gather}\label{finalham}
 \hat{\mathcal{H}} = \hat{S}^{-1}\hat{H}\hat{S} = 
\frac{(\hat{\mathbf{P}}-\hat{\mathbf{P}}_B)^2}{2m} + n_0 V(\mathbf{0}) + \sum_{\mathbf{k\neq \mathbf{0}}} \omega_{\mathbf{k}} \hat \beta^\dagger_{\mathbf{k}} \hat \beta_{\mathbf{k}}  + \sqrt{n_0} \sum_{\mathbf{q}} \sqrt{\epsilon_\mathbf{q}/\omega_\mathbf{q}} V(\mathbf{q}) \left( \hat \beta_\mathbf{q}^\dagger + \hat \beta_{\mathbf{-q}} \right)\\ \nonumber
         +  \sum_{\mathbf{k}\mathbf{k'}}  V(\mathbf{k}-\mathbf{k}') (u_{\mathbf{k'}} u_{\mathbf{k}} + v_{\mathbf{k'}} v_{\mathbf{k}}) \hat \beta^\dagger_\mathbf{k} \hat \beta_{\mathbf{k'}} 
         - \sum_{\mathbf{k}\mathbf{k'}}  V(\mathbf{k}'+\mathbf{k})  (v_{\mathbf{k'}} u_\mathbf{k}+ u_\mathbf{k'}v_{\mathbf{k}}) (\hat \beta_{\mathbf{k}}^\dagger \hat \beta_\mathbf{k'}^\dagger + \hat \beta_{-\mathbf{k}} \hat \beta_\mathbf{-k'} ).
\end{gather}

From the ansatz $|\Psi(t)\rangle = \exp(-i \phi(t))\hat{ \mathcal{O}} |\Psi(0)\rangle$ we derive the equations of motion applying a time-dependent variational principle from the Euler-Lagrange equations
\begin{gather}
\frac{d}{dt} \frac{\partial L}{\partial \dot{f}} - \frac{\partial L}{\partial f} = 0, 
\end{gather}
where $\hat{ \mathcal{O}}=e^{\sum_\mathbf{k} (\gamma_\mathbf{k}(t) \hat{\beta}^\dagger_\mathbf{k} - \gamma^\ast_\mathbf{k}(t) \hat{\beta}_\mathbf{k})}$  and 
$ f=\{\gamma, \gamma^* \}.$ Here, the Lagrangian is given by
\begin{gather}
    L = \langle \Psi  | i \partial_t - \hat{\mathcal{H}} | \Psi \rangle=i\langle \Psi  |  \partial_t | \Psi \rangle-\langle \Psi  |  \hat{\mathcal{H}} | \Psi \rangle\label{lagrangian}.
\end{gather}
After some algebra, we obtain the equations of motion for $\phi(t)$ and $\gamma(t)$ given in the main text (Eqs. [4] and [5]).
%
\subsection{Numerical procedure: Variational Ansatz}

We solve the equations of motions in Eqs.~[4] and [5] (main text) by discretising the $\mathbf k$-space, this allows us to re-write the equations of motions for $\gamma$ as a finite system of equation of motions, which can be written in a matrix form~\cite{Dzsotjan2019},
\begin{gather}
    i \dot{\bm{\gamma}} = A \bm{\gamma}+\mathbf{K},
\end{gather}
where $\bm{\gamma}=d/dt(\gamma_1, \ldots, \gamma_{N_{\text{max}}}, \gamma_1^\ast, \ldots,\gamma_{N_{\text{max}}}^\ast)$, $A$ is matrix that contains all the coefficients describing the system of equations, and $\mathbf{K}$ is a constant vector with the constant terms in the equations of motion. The solution is given by 
\begin{gather}
    \bm{\gamma}=\exp(-iAt)\mathbf{d}-A^{-1}\mathbf{K}, 
\end{gather}
where $\mathbf{d}$ is a vector that is determined from the initial condition. In our case $\bm{\gamma}(0)=\mathbf{0}$ and thus $\mathbf{d}=A^{-1}\mathbf{K}$. Once the coefficients $\gamma_\mathbf{k}$ are obtained, we obtain the dynamics of $\phi$.\\

To numerically resolve the spectral function from the Fourier transform of the dynamical overlap, we multiply the latter by an exponential decay $S(t)\rightarrow S(t)e^{-\delta t}$, this procedure is equivalent to adding an artificial broadening to the spectral function. For our numerical calculations we take $\delta=0.5$. Finally, due to the finite size of the time interval, we correct the spectral function to maintain a positive value of $A(\omega)$~\cite{Dishan1995}.

\subsection{$T$-matrix approach and few-body phonon states}
Since the potential is frequency independent one can perform the Matsubara sum in Eq.~[7] (main text)  to arrive to the following Bethe-Salpeter equation
\begin{gather}
\label{Tm}
\Gamma(\mathbf k,\mathbf Q-\mathbf k,\mathbf k-\mathbf k',\Omega)=V(\mathbf k-\mathbf k')+\int\frac{d^3\mathbf q}{(2\pi)^3}V(\mathbf k-\mathbf q)\Pi_{\mathbf q}(\mathbf Q,\Omega) \Gamma(\mathbf q,\mathbf Q-\mathbf q,\mathbf q-\mathbf k',\Omega),
\end{gather}
that describes the interaction between a boson and an ion that scatters from states with momentum-energy  $k=(\mathbf k,\omega)$ and $Q-k=(\mathbf Q-\mathbf k,\Omega-\omega)$ into states  $k'=(\mathbf k',\omega')$ and $Q-k'=(\mathbf Q-\mathbf k',\Omega-\omega').$  Here the propagator $\Pi_{\mathbf q}(\mathbf Q,\Omega)$ is given at zero-temperature by
\begin{gather}
\Pi_{\mathbf q}(\mathbf Q,\Omega)=\frac{u_{\mathbf Q/2-\mathbf q}^2}{\omega-\epsilon^I_{\mathbf Q/2+\mathbf q}-E_{\mathbf Q/2-\mathbf q}}
\end{gather}

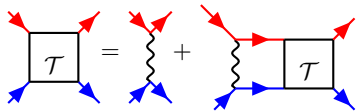
\begin{figure}[h!]
     \begin{tikzpicture}[baseline=(current bounding box.center)]
  \begin{feynman}
  \vertex (i1) ;
   \vertex [above right=.4 cm of i1] (i2);
    \vertex [ right=.65 cm of i2] (i3);
    \vertex [below right=.4 cm of i3] (i12);
    \vertex [ above=.65 cm of i3] (i4);
    \vertex [ left=.65 cm of i4] (i5);
    \vertex[above left=.4 cm of i5](i20);

    \vertex [right=.65 cm of i5] (i6);
     \vertex [ right=.65 cm of i6] (i7);
     \vertex [below=.6 cm of i7] (i8);
      \vertex [ below right=.4 cm of i3] (i9);
    \vertex [ above left=.65 cm of i7] (i11);   
     \vertex [below right=.65 cm of i8] (i10); 
     \vertex[above right=.4 cm of i4](i13);
      \vertex[above left=.5cm of i3](i14);
     \diagram* {
     (i20)--[fermion,thick,red](i5),
     (i1)--[fermion,thick,blue](i2),
      (i3)--[fermion,thick,blue](i9),
      (i2)--[thick,edge label=\(\mathcal T\)](i3)--[thick](i4)--[thick](i5)--[thick](i2),
     (i4)--[fermion,thick,red](i13),
     };
  \end{feynman}
  \end{tikzpicture}=
    \begin{tikzpicture}[baseline=(current bounding box.center)]
  \begin{feynman}
  \vertex (i1)   ;
   \vertex [above right=.4 cm of i1] (i2);
   \vertex [ above=.65 cm of i2] (i3);
    \vertex [below right=.4 cm of i2] (i4) ;
    \vertex [ above left=.4cm of i3] (i5) ;
    \vertex [ above right =.4 cm of i3] (i6) ;   
    \vertex[right=.4cm of i2](i7);
      \vertex[right=.4cm of i3](i8);
    \diagram* {
     (i1)--[fermion,thick,blue](i2),
     (i3)--[boson,thick](i2),
     (i2)--[fermion,thick,blue](i4),
     (i3)--[fermion,thick,red](i6),
     (i5)--[fermion,thick,red](i3),
           };
  \end{feynman}
  \end{tikzpicture}+
    \begin{tikzpicture}[baseline=(current bounding box.center)]
  \begin{feynman}
  \vertex (i1)   ;
   \vertex [above right=.4 cm of i1] (i2);
   \vertex [ above=.65 cm of i2] (i3);
    \vertex [below right=.4 cm of i2] (i4) ;
    \vertex [ above left=.65cm of i3] (i5) ;
    \vertex [ above right =.65 cm of i3] (i6) ;   
    \vertex[right=.65cm of i2](i7);
      \vertex[right=.65cm of i3](i8);
      \vertex[right=0.65cm of i7](i9);
      \vertex[above=0.65 cm of i9](i10);
      \vertex[above right=0.4cm of i10](i11);
      \vertex[below right=0.4cm of i9](i13);
    \diagram* {
     (i1)--[fermion,thick,blue](i2),
     (i3)--[boson,thick](i2),
     (i2)--[fermion,thick,blue](i7),
     (i3)--[fermion,thick,red](i8),
     (i5)--[fermion,thick,red](i3),
     (i9)--[fermion,thick,blue](i13),
    (i7)--[thick,edge label=\(\mathcal T\)](i9)--[thick](i10)--[thick](i8)--[thick](i7),
    (i10)--[fermion,thick,red](i11),
           };
  \end{feynman}
  \end{tikzpicture}
   \caption{Scattering matrix $\mathcal T$ for the ion-boson potential, the solid red lines correspond to the BEC Green's function, while the blue denote the ion propagator. The wavy line depicts the boson-ion interaction potential $V(\mathbf q)$.}
   \label{Tmatrices}
 \end{figure}

We focus on the low-energy scattering of the Bethe-Salpeter equation in Fig.~\ref{Tmatrices}, thus we take $k=k'=0$, and $\mathbf Q=0$ and simplify the notation for this scattering process by defining $\Gamma_0(\omega)$.  We take a quasiparticle approach to study the ionic impurity dressed by the Bogoliubov excitations, 
\begin{gather}
\Sigma(\omega)=n_B\Gamma_{0}(\omega),
\end{gather}
and calculate the dressed impurity Green's function given by
\begin{gather}
G_{I}(\mathbf p,\omega)=\frac{1}{\omega-\epsilon_{\mathbf p}^I-\Sigma(p,\omega)},
\end{gather}
here the poles of $G_I$ determine the quasiparticle energy, which can be obtained by solving self-consistently the following equation $\omega-\Sigma(\omega)=0.$ 

To determine the few-body states of many-phonons bound to the ion, we solve recursively the scattering the $T$-matrix by replacing the ion dispersion by the immediate upper polaron branch, that is, to solve the Bethe-Salpeter equation we take
\begin{gather}
\Pi_{\mathbf q}(\mathbf Q,\Omega)\rightarrow\frac{u_{\mathbf Q/2-\mathbf q}^2}{\omega-E^{P,i}_{\mathbf Q/2+\mathbf q}-E_{\mathbf Q/2-\mathbf q}},
\label{pairB}
\end{gather}
where $E^{P,i}$ denotes the polaron energy of the $i$-th upper branch.

\subsection{Molecular radius}
Fig.~\ref{weakdressingcloud2} shows that the size of the highest molecular state (the repulsive polaron with \emph{one} Bogoliubov phonon bound) decreases as it becomes increasingly bound with increasing $1/k_na$, as expected. For identical binding energy, the different molecular states will have the same radius. We can therefore calculate the gas factor for each molecular state and estimate at which interaction the Bogoliubov theory is no longer valid. 
\begin{figure}[ht]
    \includegraphics[width=0.35\columnwidth]{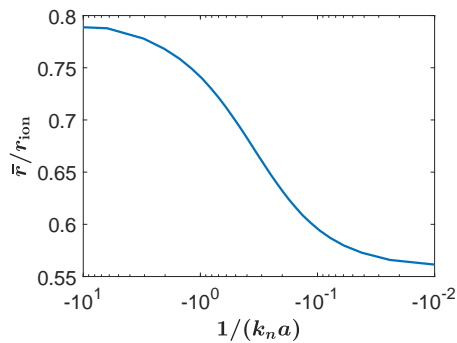}
    \caption{The spatial size of the highest molecular state as function of $1/k_n a$. As the interaction is increased, the molecular size decreases.} 
    \label{weakdressingcloud2}
\end{figure}

\end{widetext}

\end{document}